\RequirePackage{arydshln}
\documentclass[floatfix,aps,twocolumn,nofootinbib,superscriptaddress,preprintnumbers,pra,10pt]{revtex4-1}

\usepackage{colortbl} 
\definecolor{Gray}{gray}{0.95}
\definecolor{RGray}{gray}{0.93}
\definecolor{CGray}{gray}{0.92}

\usepackage[utf8]{inputenc}
\usepackage{float}
\usepackage{amsmath,amssymb,amsfonts,bm,bbm,slashed}
\usepackage{dsfont} 
\usepackage{hyperref}
\usepackage{graphicx}
\usepackage{enumitem}
\usepackage{mathtools}
\usepackage{bbold}
\usepackage{braket}
\usepackage{multirow}
\usepackage{soul}
\usepackage[table]{xcolor}
\usepackage[capitalise]{cleveref}

\usepackage[normalem]{ulem}
\usepackage{nicematrix}

\makeatletter

    \def\CT@@do@color{%
      \global\let\CT@do@color\relax
            \@tempdima\wd\z@
            \advance\@tempdima\@tempdimb
            \advance\@tempdima\@tempdimc
    \advance\@tempdimb\tabcolsep
    \advance\@tempdimc\tabcolsep
    \advance\@tempdima2\tabcolsep
            \kern-\@tempdimb
            \leaders\vrule
                    \hskip\@tempdima\@plus  1fill
            \kern-\@tempdimc
            \hskip-\wd\z@ \@plus -1fill }
\makeatother

\makeatletter
\g@addto@macro\bfseries{\boldmath}
\makeatother

\newcommand{\gsim}{\lower.7ex\hbox{$\;\stackrel{\textstyle>}{\sim}\;$}}
\newcommand{\lsim}{\lower.7ex\hbox{$\;\stackrel{\textstyle<}{\sim}\;$}}

\newcommand{\cC}{\mathcal{C}}

\newcommand{\cB}{\mathcal{B}}
\newcommand{\no}{\nonumber}

\newcommand{\xZ}{x_{\Zp}}
\newcommand{\xG}{x_{G^\prime}}
\newcommand{\mU}{m_{U_1}}

\newcommand{\Gp}{G^\prime}
\newcommand{\Zp}{Z^\prime}
\newcommand{\eps}{\epsilon}

\allowdisplaybreaks

\begin{document}

\preprint{ZU-TH-25/22}

 \title{Flavor Non-universal Vector Leptoquark 
 Imprints in $K\to \pi \nu\bar \nu$
\\ and $\Delta F = 2$ Transitions}
 
\author{\`{O}scar L. Crosas}
\email{olaracrosas@student.ethz.ch}
\affiliation{Physik-Institut, Universit\"at Z\"urich, CH-8057 Z\"urich, Switzerland}
\author{Gino Isidori}
\email{isidori@physik.uzh.ch}
\affiliation{Physik-Institut, Universit\"at Z\"urich, CH-8057 Z\"urich, Switzerland}
\author{Javier M. Lizana}
\email{jlizana@physik.uzh.ch}
\affiliation{Physik-Institut, Universit\"at Z\"urich, CH-8057 Z\"urich, Switzerland}
\author{Nud{\v z}eim Selimovi{\'c}}
\email{nudzeim@physik.uzh.ch}
\affiliation{Physik-Institut, Universit\"at Z\"urich, CH-8057 Z\"urich, Switzerland}
\author{Ben A. Stefanek}
\email{bestef@physik.uzh.ch}
\affiliation{Physik-Institut, Universit\"at Z\"urich, CH-8057 Z\"urich, Switzerland}

\begin{abstract}
\vspace{5mm}
We analyze $K\to \pi \nu\bar \nu$ rates in a model with a TeV-scale 
leptoquark addressing $B$-meson anomalies, based on the flavor non-universal 4321 gauge group featuring  third-generation quark-lepton unification. 
We show that, together with the 
tight bounds imposed by $\Delta F = 2$ amplitudes,  the present measurement of 
$\cB(K^+ \to \pi^+ \nu \bar\nu)$ already provides a non-trivial 
constraint on the model parameter space. 
In the minimal version of the model, the deviations from the Standard  Model in
$\cB(K^+ \to \pi^+ \nu \bar\nu)$ are predicted to be in 
close correlation with non-standard effects in the Lepton Flavor Universality ratios $R_D$ and
 $R_{D^*}$. With the help of future data, these correlations can provide a decisive test of the model.
\vspace{3mm}
\end{abstract}

\maketitle

\section{Introduction}\label{sec:intro}

The anomalies in
neutral-current~\cite{LHCb:2014vgu,LHCb:2017avl,LHCb:2019hip,LHCb:2021trn} and charged-current~\cite{BaBar:2012obs,BaBar:2013mob,Belle:2015qfa,LHCb:2015gmp,LHCb:2017smo,LHCb:2017rln}  $B$-meson decays 
have triggered a renewed interest in precision tests of semi-leptonic processes, and 
in particular in transitions involving third-generation fermions. In this general context,
$K\to \pi \nu\bar \nu$ decays play a very special role. First, these processes are among the very 
few flavor-changing transitions involving light quarks $(s \to d)$  that we are able to compute precisely within the Standard Model (SM)
(see~\cite{Buras:2015qea} for state-of-the art predictions). Second, they are extremely suppressed 
within the SM due to its approximate accidental symmetries, making them a particularly sensitive probe of possible new physics (NP). Third, these processes uniquely probe the couplings of light quarks to third-generation 
leptons via tau neutrinos in the final state. Last but not least, the NA62 experiment at CERN is now sensitive enough to test the SM prediction for the branching ratio of the charged kaon decay mode, $\cB(K^+ \to \pi^+ \nu \bar\nu)$~\cite{NA62:2021zjw}.

The connection between $K \to \pi \nu \nu$ and the $B$ anomalies has been explored in a series of recent  papers,
both adopting a general EFT approach~\cite{Bordone:2017lsy,Descotes-Genon:2020buf}, as well as in specific new physics models
\cite{Fajfer:2018bfj,Crivellin:2018yvo,Marzocca:2021miv}. From these previous studies, it emerges quite 
clearly that an $O(10\%)$ deviation from the SM in the Lepton Flavor Universality (LFU) ratios $R_{D^{(*)}}$ naturally implies  
$O(1)$ deviations from the SM in $\cB(K^+ \to \pi^+ \nu \bar\nu)$. However, the precise size of the effect is rather model dependent. 

In this paper, we address the precise connection between $R_{D^{(*)}}$ and  $\cB(K^+  \to \pi^+ \nu \bar\nu)$  in the framework of a 
particularly motivated class of SM extensions. We focus on models where the $B$ anomalies are described by the exchange of a TeV-scale vector leptoquark (LQ)~\cite{Alonso:2015sja,Calibbi:2015kma,Barbieri:2015yvd,Bhattacharya:2016mcc,Buttazzo:2017ixm}
arising from the spontaneous symmetry breaking of the flavor non-universal gauge group 
$SU(4)_h \times SU(3)_l \times SU(2)_L \times U(1)_X$  down to the SM~\cite{DiLuzio:2017vat,Bordone:2017bld,Greljo:2018tuh,Fuentes-Martin:2019ign,Fuentes-Martin:2020hvc}.  In these so-called non-universal 4321 models, 
color is embedded as the diagonal subgroup of $SU(4)_h \times SU(3)_l$~\cite{DiLuzio:2017vat}, where the labels $h$ (heavy) and $l$ (light) indicate the flavor non-universal charge assignment of the SM fermions under this part of the gauge group. This setup is particularly appealing as it results in quark-lepton unification a l\`a Pati-Salam~\cite{Pati:1974yy} for  third-generation fermions, an accidental approximate $U(2)^5$ flavor symmetry, and a $U_1$ vector LQ that is automatically dominantly coupled to the third family~\cite{Bordone:2017bld,Greljo:2018tuh,Fuentes-Martin:2019ign,Fuentes-Martin:2020hvc}.
As recently shown in~\cite{Fuentes-Martin:2022xnb,Fuentes-Martin:2020pww}  
this setup can arise from a higher dimensional (or multi-scale) construction addressing both the origin of the SM flavor 
hierarchies~\cite{Panico:2016ull,Bordone:2017bld,Allwicher:2020esa,Barbieri:2021wrc}
 as well as the electroweak (EW) hierarchy problem~\cite{Barbieri:2017tuq,Fuentes-Martin:2020bnh}, 
 with an intrinsic theoretical interest that goes beyond the $B$ anomalies. 

This ambitious class of models contains a sizeable number of new fields compared to the SM. In particular, 
in addition to the LQ, two neutral massive vectors (an octet and a singlet of color), as well as one or more families of vector-like (VL) fermions are necessarily present: these exotic (heavy) fermions are required in order to generate mixing between the light and third generations, which is needed both in the Yukawa sector as well as in the couplings to the new massive gauge fields. Despite the presence of new exotic fields, a large fraction of the model parameters are fixed by observable quantities. These include the $B$ anomalies, a series of tight constraints from various low-energy precision tests,  
and the elements of the Cabibbo-Kobayashi-Maskawa (CKM) matrix, which are now calculable in the model in terms of more fundamental parameters. This results in a highly predictive framework, at least in the minimal version of the model with only one family of VL fermions, which is what we consider in this paper. As we show, the tight constraints from $\Delta F=2$ observables (in particular $\epsilon_K$ and 
CP violation in $D$--$\bar D$ mixing) imply a well-defined structure for the couplings of the light quarks to the new heavy states that, in turn, leads to a clear correlation between $R_{D^{(*)}}$ and  $\cB(K^+ \to \pi^+ \nu \bar\nu)$.

The paper is organised as follows. In Section~\ref{sec:model} we introduce the model.
In Section~\ref{sec:deltaF2}, we analyze $\Delta F=2$ observables and determine the range of 
the parameters controlling flavor mixing in the light-quark sector. Using these results, in
Section~\ref{sec:Kpinunubar} we analyze the predictions for  $\cB(K^+ \to \pi^+ \nu \bar\nu)$.
The results are summarized in the conclusions. 
The appendices contain: 
\ref{app:appendixA})
a detailed analysis of the flavor structure of the model,
including the computation of the CKM matrix;
\ref{app:appendixC})
a discussion about the inclusion of additional 
VL fermions charged under $SU(3)_l$;
\ref{app:appendixD}) expressions relevant to estimate $R_{D^{(*)}}$;
\ref{app:appendixB})
complete results for the Wilson coefficients 
(both  tree- and one-loop contributions)
for the effective operators 
relevant to $\Delta F=2$ and $K\to \pi\nu\bar{\nu}$ observables.

\section{The Model}\label{sec:model}
The model we consider is based on the gauge group $\mathcal{G}_{4321} = SU(4)_h \times SU(3)_l \times SU(2)_L \times U(1)_X$, 
where color is the diagonal subgroup of $SU(4)_h \times SU(3)_l$ and the flavor-universal group $SU(2)_L$ acts as in the SM.
The $U(1)_X$ group acts as the SM hypercharge on the first two families, while for the third family 
$Y=X+\sqrt{2/3}\,T_4^{15}$, where $T_4^{15}=\frac{1}{2\sqrt{6}}\mathrm{diag}(1,1,1,-3)$ is a generator
of $SU(4)_h$. The matter content  is reported in Table~\ref{tab:fieldcontent}. In the absence of mixing, 
the three  $SU(4)_h$ multiplets $\psi_L$ and $\psi_R^\pm$ can be identified with the 
SM third-generation fermions (with the addition of a right-handed neutrino). The only exotic fermion is a single
VL fermion ($\chi_{L,R}$), where both chiralities have the same gauge quantum numbers as the chiral field $\psi_L$. 

The massive gauge bosons resulting from the spontaneous symmetry breaking (SSB) $4321\to$~SM  
transform under the SM gauge group as $U_1\sim(\bf{3},\bf{1},2/3)$, $G^\prime\sim(\bf{8},\bf{1},0)$ and $Z^\prime\sim(\bf{1},\bf{1},0)$. 
This SSB occurs at the TeV scale via the vacuum expectation values of two scalar fields transforming in the anti-fundamental of $SU(4)$, $\Omega_1$ and  $\Omega_3$, which are singlets and triplets under $SU(3)_l$, respectively. Let us denote the four gauge couplings of $\mathcal{G}_{4321}$ as $g_i$ ($i=1,\ldots4$).\footnote{In terms of the $g_i$,  the SM gauge couplings for color and hypercharge 
are $g_c = g_3 \cos\theta_3$ and   $g_Y =g_1 \cos\theta_1$.}
In terms of the mixing angles $\tan^2 \theta_{3} = g_{3}^2/g_4^2$ and $\tan^2\theta_{1} = 2g_{1}^2/3g_4^2$, the gauge boson masses read
\begin{align}\label{eq:SU4vectorMasses}
\mU = \frac{g_{4} f_{U_1}}{2} \,, \hspace{10mm} m_{Z',G'} = \frac{\mU}{\cos\theta_{1,3}} \frac{f_{Z',G'}}{f_{U_1}} \,,
\end{align}
where $f_{U_1}^2 = \omega_1^2 + \omega_3^2$, $f_{Z'}^2 =3 \omega_1^2 /2+ \omega_3^2/2$, $f_{G'}^2 = 2\omega_{3}^2$,
and $\omega_{1,3}$ are the $\Omega_{1,3}$ vacuum expectation values. Note that in the custodial $SU(4)_V$ limit $\omega_3=\omega_1$, all masses are approximately degenerate in the phenomenological limit $g_{1,3} \ll g_4$. More details on the model can be found in~\cite{Fuentes-Martin:2020hvc}.

\begin{table}[t]
\begin{center}
\begin{tabular}{|c|c|c|c|c|}
\hline
Field & $SU(4)_h$ & $SU(3)_l$ & $SU(2)_L$ & $U(1)_{X}$ \\
\hline
\hline
$q^i_L$ & $\mathbf{1}$ & $\mathbf{3}$ & $\mathbf{2}$ & $1/6$ \\
$u^i_R$ & $\mathbf{1}$ & $\mathbf{3}$ & $\mathbf{1}$ & $2/3$  \\
$d^i_R$ & $\mathbf{1}$ & $\mathbf{3}$ & $\mathbf{1}$ & $-1/3$  \\
$\ell^i_L$ & $\mathbf{1}$ & $\mathbf{1}$ & $\mathbf{2}$ & $-1/2$ \\
$e^i_R$ & $\mathbf{1}$ & $\mathbf{1}$ & $\mathbf{1}$ & $-1$ \\ 
$\psi_L$ & $\mathbf{4}$ & $\mathbf{1}$ & $\mathbf{2}$ & $0$ \\ 
$\psi_R^{\pm}$ & $\mathbf{4}$ & $\mathbf{1}$ & $\mathbf{1}$ & $\pm1/2$ \\  \rowcolor{RGray}
$\chi_{L,R}$ & $\mathbf{4}$ & $\mathbf{1}$ & $\mathbf{2}$ & 0  \\ 
\hline
\hline
$H$ & $\mathbf{1}$ & $\mathbf{1}$ & $\mathbf{2}$ & 1/2  \\     \rowcolor{RGray}
$\Omega_1$ & $\mathbf{\bar 4}$ & $\mathbf{1}$ & $\mathbf{1}$ & $-1/2$  \\ \rowcolor{RGray}
$\Omega_3$ & $\mathbf{\bar 4}$ & $\mathbf{3}$ & $\mathbf{1}$ & $1/6$  \\
\hline
\end{tabular}
\end{center}
\caption{Minimal matter-field content of the model ($i=1,2$). Fields added to the SM matter content are shown in grey.
}
\label{tab:fieldcontent}
\end{table}

After the TeV-scale SSB, a mass mixing is induced between chiral and VL fermions. Let us define $\Psi_{q}^{\prime \intercal} = (q_{L}^{\prime 1} \; q_{L}^{\prime 2} \; q_L^{\prime 3} \;  Q^{\prime}_L)$ and similarly for the left-handed (LH) leptons. 
Here, $Q_L$ denotes the LH quark component of the VL field and the prime superscript indicates we are in the interaction (or gauge-eigenstate) basis and not 
in the mass-eigenstate basis. The mass mixing can be written as 
\begin{equation}
-\mathcal{L} \supset \bar \Psi_{q}^{\prime} \, \mathbf{M}_{q}  Q_R + \bar \Psi_{\ell}^{\prime} \, \mathbf{M}_{\ell}  L_R + \textrm{h.c.} \,,
\label{eq:VLmass}
\end{equation}
and, without loss of generality, we can decompose the mass matrices as
\begin{align}
\mathbf{M}_{q} &= \mathbf{W}_q^{\dagger} \, {\bf O}_q^{\dagger} \,\, (0\;\, 0\;\, 0\;\, m_Q)^{\intercal} \nonumber \,, \\
\mathbf{M}_{\ell} &= \mathbf{W}_\ell^{\dagger}\, {\bf O}_\ell^{\dagger}\,\, (0\;\, 0\;\, 0\;\, m_L)^{\intercal} \,,
\end{align}
where $m_Q$ and $m_L$ are TeV-scale masses. The $4\times 4$ unitary rotation matrices can be parameterized as
\begin{align}
{\bf O}_{q,\ell} = 
\begin{pmatrix}
1 & 0 & 0 & 0 \\
0 & c_{q,\ell}  & 0 & -s_{q,\ell}\\
0 & 0 & 1 & 0  \\
0 & s_{q,\ell} & 0 & c_{q,\ell}
\end{pmatrix}   \,, \hspace{3mm}
{\bf W}_{q,\ell} = 
\begin{pmatrix}
\mathbb{1}_{2\times 2} & 0_{2\times 2}  \\
0_{2\times 2} & W_{q,\ell}  \\
\end{pmatrix}   \,,
\end{align}
where $W_{q,\ell}$ are $2\times2$ unitary matrices and throughout the paper we adopt the notation 
$s_{q,\ell} \equiv \sin \theta_{q,\ell} $ and  $c_{q,\ell}  \equiv \cos \theta_{q,\ell} $. 
We can move to the VL fermion mass basis 
(where three chiral fermions species remain massless until EW symmetry breaking) 
by redefining the LH quark and lepton fields as $\Psi_{q,\ell}^{\prime} \rightarrow \mathbf{W}_{q,\ell}^{\dagger} \, {\bf O}_{q,\ell}^{\dagger}  \Psi_{q,\ell}$. In this basis, the interactions of LH fermions with the massive gauge bosons 
read
\begin{align}\label{eq:ULag}
\mathcal{L}_{U_1}&\supset\frac{g_4}{\sqrt{2}}\, U^{\mu}_{1}   \, (\bar \Psi_q \beta_L \gamma_{\mu} \Psi_{\ell})  \,, \\
\mathcal{L}_{G^{\prime}}&\supset g_4 \frac{ g_c }{g_3}\,G^{\prime \, a}_\mu  \,(\bar \Psi_q \kappa_q\gamma^\mu\,T^a\, \Psi_q) \,, \\
\mathcal{L}_{Z^\prime} &\supset \frac{g_4}{2\sqrt{6}} \frac{g_Y }{g_1}\,Z_\mu^\prime \left[(\bar \Psi_q \xi_q \gamma^\mu \Psi_q)-3(\bar \Psi_\ell \xi_\ell \gamma^\mu \Psi_\ell) \right] \,,
\label{eq:ULagEnd}
\end{align}
where, defining $\eps_{3} = g_3^2/g_4^2$, $\eps_{1} = 2g_1^2/3g_4^2$, the couplings are given as
\begin{align}
\kappa_q &= {\bf O}_{q} \kappa_q^\prime {\bf O}_{q}^{\dagger} \,, \hspace{8mm} \kappa_q^\prime=\mathrm{diag}(-\eps_3,-\eps_3,1,1)\,, \\
\xi_q &= {\bf O}_{q} \, \xi_{q}^\prime {\bf O}_{q}^{\dagger} \,, \,  \hspace{8mm} \xi_q^\prime=\mathrm{diag}(-\eps_1,-\eps_1,1,1)\,, \\
\xi_\ell &= {\bf O}_{\ell} \, \xi_{\ell}^\prime \, {\bf O}_{\ell}^{\dagger} \,, \, \hspace{8mm} \xi_\ell^\prime=\mathrm{diag}(-\eps_1,-\eps_1,1,1)\,,
\end{align}
and
\begin{equation}
\beta_L = {\bf O}_{q} {\bf W} \beta_L^\prime {\bf O}_{\ell}^{\dagger} \,, \hspace{8mm} 
{\bf W} = 
\begin{pmatrix}
\mathbb{1}_{2\times 2} & 0_{2\times 2}  \\
0_{2\times 2} & W  \\
\end{pmatrix}   \,,
\end{equation}
with $\beta_L^\prime=\mathrm{diag}(0,0,1,1)$. We give explicit formulas for the 4321 gauge boson couplings in Appendix~\ref{app:appendixGB}.

The matrix $W = W_q W_{\ell}^{\dagger}$ can always be written as a real, orthogonal $2\times 2$ matrix by re-phasing $q_L^3$, $\ell_L^3$, and the LH VL fermions.  We parameterize it by an angle $\theta_\chi$:
\begin{equation}
W = \begin{pmatrix}
c_{\chi} & s_{\chi}  \\
-s_{\chi} & c_{\chi}  \\
\end{pmatrix}   \,.
\end{equation}
A non-trivial $W$-matrix (i.e.~$s_\chi\not=0$) arises from the difference between the quark and lepton mass matrices in (\ref{eq:VLmass}), in the sub-sector of the $SU(4)$-charged fermions~\cite{Fuentes-Martin:2020hvc}.
It can be generated by including a scalar field $\Omega_{15}$, transforming as a $({\bf 15},{\bf 1},{\bf 1},0)$ of $\mathcal{G}_{4321}$, which takes a vacuum expectation value $\omega_{15}$ along the $T_4^{15}$ generator. This would also shift the mass of the $U_1$, 
modifying the $f^2_{U_1}$ appearing in \eqref{eq:SU4vectorMasses}
to $f_{U_1}^2=\omega_1^2+\omega_3^2+4/3 \omega_{15}^2$.
Note that the re-phasing that makes $W$ real  
results in a phase appearing in the LQ coupling to the right-handed (RH) VL fermions. However, this phase has no impact on the low-energy phenomenology so we do not discuss it further.

\subsection{Yukawa Couplings}\label{sec:YukCouplings}
We are now ready to analyze the structure of the Yukawa couplings, i.e.~the couplings of the four families of $SU(2)_L$
charged fields, the three right-handed chiral fermions, and the Higgs field. Defining $\Psi_{u}^\intercal = (u_{R}^1 \; u_{R}^2 \; u_R^3)$, $\Psi_{d}^\intercal = (d_{R}^1 \; d_{R}^2 \; d_R^3)$, $\Psi_{e}^\intercal = (e_{R}^1 \; e_{R}^2 \; e_R^3)$, the Yukawa couplings in the VL fermion mass basis can be written as
\begin{align}
-\mathcal{L} & \supset \bar \Psi_{q} {\bf Y}_u \tilde{H} \Psi_{u} +  \bar \Psi_{q} {\bf Y}_d H \Psi_{d}  + \bar \Psi_{\ell} {\bf Y}_e H \Psi_{e}  + \textrm{h.c.} \,,
\end{align}
where ${\bf Y}_{u,d,e}$ are $4\times 3$ matrices that can be decomposed as
\begin{align}
{\bf Y}_u  = {\bf O}_{q} \,
&\begin{pNiceMatrix}
U_{u}^{\dagger}   \hat Y_u  & \vec{n}_u  \\
0  & y_t   \\ \hline
0  &y_t x_{t}  e^{i \varphi_{t}} 
\end{pNiceMatrix}
\,, \label{eq:Yu} \\
{\bf Y}_d  = {\bf O}_{q} \,
&\begin{pNiceMatrix}
U_{d}^{\dagger}   \hat Y_d  & \vec{n}_d  \\
0  & y_b \\ \hline
0  & y_b x_b  e^{i \varphi_{b}} 
\end{pNiceMatrix}
\,, \label{eq:Yd} \\
{\bf Y}_e  = {\bf O}_\ell \, {\bf W}^{\dagger} \,
&\begin{pNiceMatrix}
U_{e}^{\dagger}   \hat Y_e  & \vec{n}_e  \\
0  & y_b  \\ \hline 
0  & y_b x_b  e^{i \varphi_{b}}
\end{pNiceMatrix} \,.
\label{eq:Ye}
\end{align}
The overall factors ${\bf O}_{q}$ and ${\bf O}_\ell \, {\bf W}^{\dagger}$ are the results of the rotation from the interaction basis 
to the VL fermion mass basis, and take into account  a global re-definition of both quark and lepton Yukawa  
of the type $ {\bf Y}' \rightarrow {\bf W}_q^{\dagger} {\bf Y}$, such that the full $\bf W$ matrix appears only in the lepton sector. 
As far as the other terms appearing in (\ref{eq:Yu})--(\ref{eq:Ye}) are concerned:
\begin{itemize}
\item{}
The hatted Yukawas $\hat Y_{u,d,e}$ are real, diagonal $2\times 2$ matrices, and  $U_{u,d,e}$ are $2\times2$ unitary matrices with 
a non-trivial phase each (after field redefinitions). These terms, which control the $2\times 2$ light-family masses (hence the smallest eigenvalues in ${\bf Y}_{u,d,e}$) 
are SM-type Yukawa couplings among the chiral fields not charged under $SU(4)$. 
\item{}
The entries $y_{t,b}$ and $x_{t,b}\, e^{i\varphi_{t,b}}$ control the couplings among the $SU(4)$ charged states, 
i.e.~the third-generation Yukawa couplings ($y_{t,b}$) and the Yukawa couplings of the VL fermion to the RH third-generation chiral fermions ($x_{t,b}$). 
  Note that the $W$-matrix leads to an effective splitting of bottom and tau Yukawas of the type
\begin{equation}
y_\tau = y_b (c_{\chi} -  x_b e^{i\varphi_b} s_{\chi}) \,.
\end{equation}
On general grounds, additional sub-leading $SU(4)$-breaking corrections are expected from integrating-out heavier fields.
\item{}
The $2\times 1$ vectors $\vec{n}_{u,d}$ and $\vec{n}_{e}$ are spurions parameterizing the $SU(4)$ breaking in the Yukawa sector
or, equivalently, the breaking of the $U(2)_q$ and $U(2)_\ell$ flavor symmetries preventing heavy $\leftrightarrow$ light mixing amongst chiral fermions~\cite{Barbieri:2015yvd,Buttazzo:2017ixm}. Given the field content in Table~\ref{tab:fieldcontent}, these terms are forbidden at the renormalizable level.
In the limit $\vec{n}_{u,d}\to0$, the only source of $U(2)_{q}$ breaking (besides the smaller light family Yukawa couplings) comes from $s_q$.
As we shall show (see also~\cite{Cornella:2019hct}), 
this case is severely constrained by $\Delta F = 2$ observables.
This is why we consider a small but non-negligible  $\vec{n}_{u}$ 
of the form $\vec{n}_{u} = (-y_t \epsilon_U e^{i \varphi_U}\; 0)^{\intercal}$, with $\epsilon_U =O(|V_{ub}|)$. As we show in Appendix~\ref{app:appendixC}, this structure can be easily generated by integrating out a heavy vector-like quark $U_{L,R}$, charged under $SU(3)_l$, that mixes with $t_R$ after 4321 symmetry breaking. Since the mixing required to generate $V_{ub}$ is small, these new states could naturally be part of deeper UV dynamics in the 10-100 TeV range as occurs in~\cite{Fuentes-Martin:2022xnb}. 
\end{itemize}

Since $U(3)_{q,\ell}$ rotations leave the VL mass invariant, the $3\times3$ upper sub-block of the Yukawas can be diagonalized by bi-unitary rotations of the form $L_u  {\bf Y}_u R_u^{\dagger} = \hat {\bf Y}_u$, $L_d  {\bf Y}_d R_d^{\dagger} = \hat {\bf Y}_d$, and $L_e  {\bf Y}_e R_e^{\dagger} = \hat {\bf Y}_e$, where $\hat {\bf Y}_{u,d,e}$ are diagonal in the $3\times3$ sub-block above the solid line and unchanged below it. In terms of these rotation matrices, the quark Yukawas can be written as
\begin{align}
\begin{aligned}
-\mathcal{L} & \supset \bar \Psi_{q} L_u^{\dagger} \hat {\bf Y}_u R_u \tilde{H}  \Psi_{u} +  \bar \Psi_{q}  L_d^{\dagger} \hat{\bf Y}_d R_d H \Psi_{d}   + \textrm{h.c.} \,.
\end{aligned}
\end{align}
The lepton Yukawas follow similarly, where acceptable neutrino masses can be achieved via the inverse-seesaw mechanism~\cite{Greljo:2018tuh}.  With these definitions, the CKM matrix is defined as $V_{\rm CKM} = L_u L_d^{\dagger}$. Explicit  
formulas for these rotation matrices, in terms of the parameters 
appearing in (\ref{eq:Yu})--(\ref{eq:Ye}) (including those {\em hidden} in  $\hat Y_{u,d,e}$ and $\hat U_{u,d,e}$) are given in~\cref{app:appendixA}.

\section{$\Delta F = 2$ Observables}\label{sec:deltaF2}
Before EW symmetry breaking, we define the effective operators
\begin{align}\label{eq:4qL_SMEFT}
\mathcal{O}_{q q}^{(1)ijkl}&=(\bar q_L^{i}\gamma_\mu q_L^{j})(\bar q_L^{k}\gamma_\mu q_L^{l})\,,\\
\mathcal{O}_{q q}^{(3)ijkl}&=(\bar q_L^{i}\gamma_\mu\tau^I q_L^{j})(\bar q_L^{k}\gamma_\mu\tau^I q_L^{l})\,,
\end{align}
where $i,j,k,l=1,2,3$ are flavor indices, $\tau^I$ the Pauli matrices, and color is contracted inside the parenthesis.
The part of the effective Lagrangian relevant for $\Delta F=2$ transitions can then be written as
\begin{align}\label{eq:DeltaQ2_SMEFT}
    \mathcal{L}_{\rm eff}  \supset -\frac{4G_U}{\sqrt{2}}\sum_{a=1,3} \mathcal{C}_{q q}^{(a)ijkl}  \mathcal{O}_{q q}^{(a)ijkl}+{\rm h.c.}\,,
\end{align}
with the sum over the flavor indices. Here, $G_U=G_F\,C_U$, where we have defined
\begin{equation}
C_U = \frac{g_4^2 v^2}{4 \mU^2} \,,
\end{equation} that controls the strength of the purely third-generation contact interactions generated by integrating out the $U_1$ LQ.
The flavor-conserving Wilson coefficients, $\mathcal{C}_{q q}^{(a)iijj}$ and $\mathcal{C}_{q q}^{(a)ijji}$, receive contributions from tree-level $G'$ and $Z'$  exchange, as well as from the $U_1$ vector LQ at one loop if $i\land j = 2 \vee 3$. The flavor-violating terms, $\mathcal{C}^{(a)ijkl}_{qq}$ with $\{i,k \}\neq \{j,l\}$, receive only loop contributions. At the matching scale, the contribution to the various Wilson coefficients of four-quark operators are reported in~\cref{tab:4QWCs} of~\cref{app:appendixB}.

\subsection{$B_{s(d)}-\bar{B}_{s(d)}$ mixing}

We start by briefly summarizing the case of 
$B_{s(d)}-\bar{B}_{s(d)}$, which has been 
extensively discussed in the previous literature (see e.g.~\cite{DiLuzio:2018zxy,DiLuzio:2019jyq,Cornella:2019hct,Cornella:2021sby}). In this case we are interested in matching to the following effective operators after EW symmetry breaking
\begin{align}
    \mathcal{L} &\supset -C_{B_s}^1(\bar{s}_L\gamma_\mu b_L)^2  -C_{B_d}^1(\bar{d}_L\gamma_\mu b_L)^2\,.
\end{align}
Performing the left-handed diagonalization of Yukawa matrices, the  complete expressions for 
$C_{B_{s(d)}}^1$ are
\begin{align}
        C_{B_s}^1 &= \frac{4G_U}{\sqrt{2}}\sum_{a=1,3} L_d^{si}L_d^{sk} \, \mathcal{C}_{q q}^{(a)ijkl} (L_d^{bj})^{*}(L_d^{bl})^{*}\,,\\
        C_{B_d}^1 &= \frac{4G_U}{\sqrt{2}}\sum_{a=1,3} L_d^{di}L_d^{dk} \, \mathcal{C}_{q q}^{(a)ijkl} (L_d^{bj})^{*}(L_d^{bl})^{*}\,.
\end{align}
Expanding the $L_d$ matrices to the first non-trivial order, and taking into account the relevant Wilson coefficients reported in Table \ref{tab:4QWCs},
both at tree- and one-loop- level,
leads to 
\begin{align}
  C_{B_{s}}^1 & =  \frac{4G_U}{\sqrt{2}} e^{2 i \phi_b}\left( \rho_{q}^{\rm NLO} s_{b}^2  + e^{-2 i \varphi_b}\frac{\alpha_4}{8\pi} s_q^2 B_{qq}^{1221} \right)\label{eq:CBs1}\,, \\
  C_{B_{d}}^1 & =  s_d^2 \, e^{2i\phi_d}  \, C_{B_{s}}^1 \,.\
\end{align}
In these expressions, we defined
\begin{equation}\label{eq:NqNLO}
  \rho_{q}^{\rm NLO}  = \bigg(\frac{1}{3\xG}+\frac{1}{24\xZ}+\frac{\alpha_4}{16\pi}\bigg)\,,
\end{equation}
where $x_{V}=m_{V}^2/\mU^2$ and $\alpha_4 = g_{4}^2/(4\pi)$:
this coefficient encodes the effect of 
flavor-conserving Wilson coefficients generated by the tree-level $\Gp$ and $\Zp$ exchange, and corresponding dominant next-to-leading order (NLO) corrections (in the  $SU(4)_V$ custodial limit). The second term in $C_{B_{s}}^1$ is due to the flavor-violating Wilson coefficient $C_{qq}^{2323}$ which is generated at one-loop.

The implications of these expressions, taking 
into account the experimental bounds on $B_{s(d)}-\bar{B}_{s(d)}$ mixing
 and the preferred range of $G_U$  and $s_q$ implied 
by the $B$ anomalies, 
can be summarized as follows~\cite{Cornella:2021sby}:
\begin{itemize}
\item $s_b \lesssim 0.2 |V_{ts}|$ is required to ensure a sufficient suppression of the flavor-conserving part of the amplitude, which is dominantly induced by tree-level $G'$ exchange.
\item Given $B_{qq}^{1221} (x_L)\propto x_L = m_L^2/M_U^2$ for small $x_L$,  light vector-like leptons ($m_L \lesssim 1.5$~TeV) are required to ensure sufficient suppression of the
$B_{qq}^{1221}$ flavor-violating loop
contribution (coming dominantly from $U_1$  box diagrams). 
\end{itemize}
The mild down-alignment required in the 23 quark sector discussed in the first point is the main motivation 
behind the power-counting adopted in~\eqref{eq:PowerCountAng} to organize the parameters appearing in the Yukawa couplings.

\subsection{$K$--$\bar{K}$ and $D$--$\bar{D}$ mixing}
After EW symmetry breaking, the effective operators describing neutral meson mixing involving only light quarks are
\begin{equation}\label{eq:DDLagrangian}
    \mathcal{L} \supset -C_K^1(\bar{d}_L\gamma_\mu s_L)^2 -C_D^1(\bar{u}_L\gamma_\mu c_L)^2\,.
\end{equation}
Again performing the left-handed diagonalization of Yukawa matrices, the expressions for $C_{K,D}^1$ read 
\begin{eqnarray}
   C_K^1 &=& \frac{4G_U}{\sqrt{2}} \sum_{a=1,3} L_d^{di}L_d^{dk} \, \mathcal{C}_{q q}^{(a)ijkl} (L_d^{sj})^{*}(L_d^{sl})^{*}\,,\\
  C_D^1 &=& \frac{4G_U}{\sqrt{2}} \sum_{a=1,3} L_u^{ui}L_u^{uk} \, \mathcal{C}_{q q}^{(a)ijkl} (L_u^{cj})^{*}(L_u^{cl})^{*}\,.
\label{eq:CD1_full}
\end{eqnarray}
In both cases the sum is dominated by the flavor conserving Wilson coefficients,
which receive tree-level contributions from $\Gp$ and $\Zp$ exchange. Given the 
power-counting 
in~\eqref{eq:PowerCountAng}
(i.e.~working in the limit $s_b\to 0$) the leading contributions are
\begin{eqnarray}
C_K^1 &=&\frac{4G_U}{\sqrt{2}}   \rho_q^{\rm NLO}   s_q^4  \left(V_{us} + c_d s_u e^{i\phi_u} \right)^2 c_d^2/c_u^2 \,,
\label{eq:CK1_leading} \\
C_D^1 &=&\frac{4G_U}{\sqrt{2}}   \rho_q^{\rm NLO}   \left(c_us_us_q^2 e^{i\phi_u}-e^{-i\delta}|V_{ub}||V_{cb}|\right)^2\,, \quad 
\end{eqnarray}
where in $C_K^1$ we have used the CKM relation $V_{us}  = e^{i\phi_d} c_u s_d - e^{i\phi_u} c_d s_u$ from~\cref{eq:CKMelems}.
According to data on the anomalies, $C_U$ lies in the range $0.004 \leq C_U \leq 0.03$.
In the  $SU(4)_V$ custodial limit (where the heavy-vector spectrum is approximately degenerate), we have $\rho_{[q]}^{\rm NLO} \approx 0.4$. 
Due to the strong scaling of $C_{K,D}^1$ with $s_q$, we focus on two benchmark points (BPs) near the upper end of the $C_U$ range, 
in order to maximize the contribution to $\delta R_{D^{(*)}}$ (defined in \eqref{eq:dRD} and~\eqref{eq:dRDstar}),
\begin{align}
{\rm BP1:} \hspace{3mm} & C_U = 0.015\,,  \\
{\rm BP2:} \hspace{3mm} & C_U = 0.030\,.
\end{align}
In both cases we assume $\xG = \xZ = 1$ unless otherwise specified.

\begin{figure}[t]
\includegraphics[width=\columnwidth]{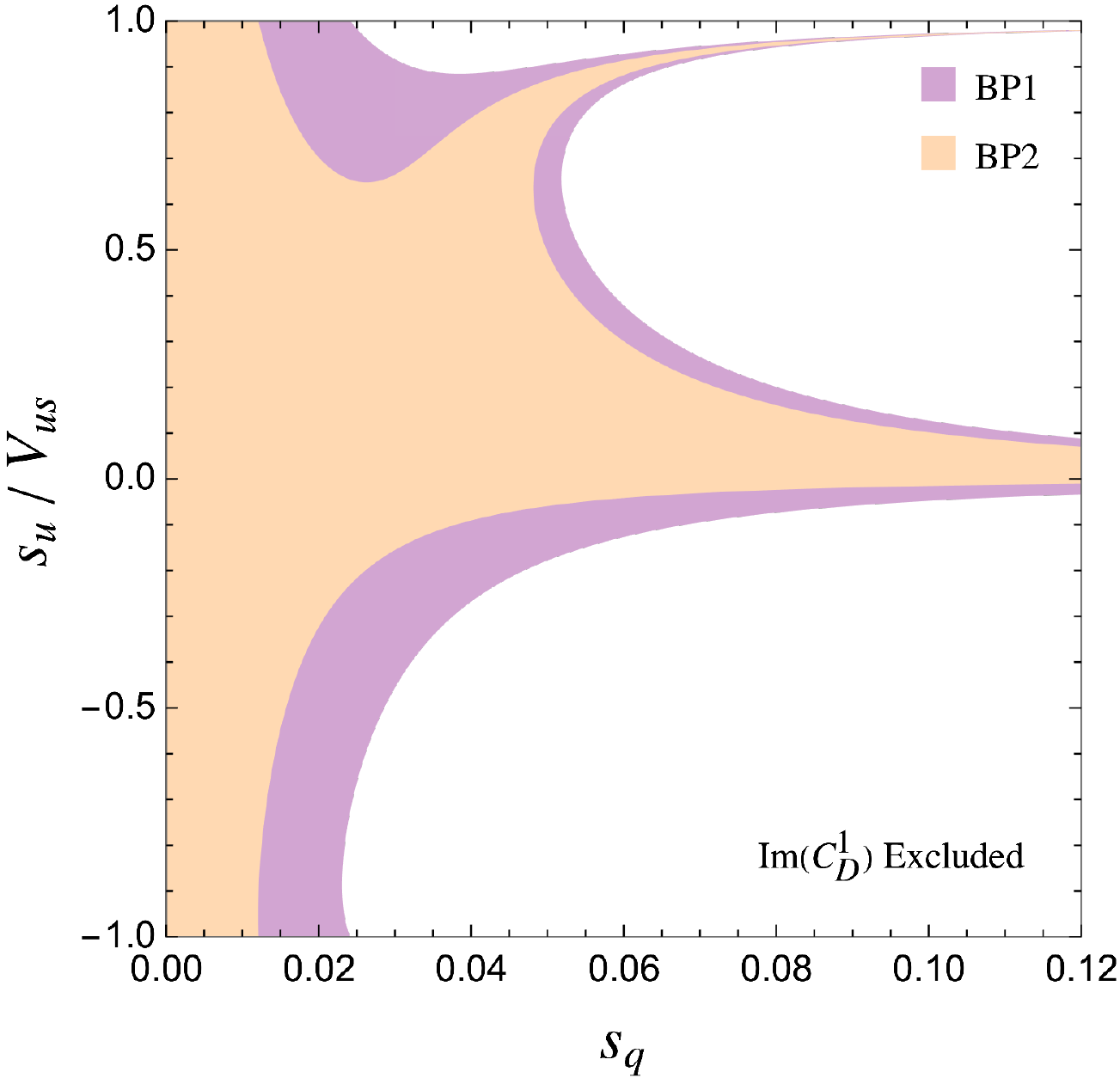}
\caption{Constraints from CP violation 
in  $D$--$\bar{D}$ mixing on the light quark mixing angle $s_u$
as a function of $s_q$, assuming a generic $\mathcal{O}(1)$ 
value for the phase $\Delta_{ud}$ (see \cref{app:appendixA}). The areas outside the colored regions (corresponding to the two benchmark points) are excluded by the bounds on ${\rm Im} C_D^1$ at the $95\%$ confidence level.}
\label{fig:DDbarplot}
\end{figure}

The strongest constraints are those on the imaginary parts of $C_{K,D}^1$. According to~\cite{FlavConstraints,UTfit:2007eik}
the $95\%$ confidence allowed range is
\begin{eqnarray}
{\rm Im}(C_K^1) &=& [-1.2,\,2.4]\times 10^{-9} \,\, {\rm TeV^{-2}}\,, \label{eq:ImKKbar} \\
{\rm Im}(C_D^1) &=& [-9.4,\,8.9]\times 10^{-9} \,\, {\rm TeV^{-2}}\,.
\label{eq:ImDDbar}
\end{eqnarray}
Comparing the expressions for $C_{K,D}^1$ to the bounds in (\ref{eq:ImKKbar})--(\ref{eq:ImDDbar}), 
it is clear that a precise flavor-alignment pattern is needed to satisfy the $\Delta F=2$ bounds
if  $| s_q | \gsim |V_{cb}|$, as required by the best fit of the charged-current anomalies.
Starting with $D$--$\bar D$ mixing, we see from \cref{fig:DDbarplot} that the simplest solution to pass the bound on ${\rm Im}(C_D^1)$ is a mild  \emph{up-alignment}:
the angle $s_u$,
controlling the 12 up-quark rotation hence naturally expected 
to be of $O(|V_{us}|)$, cannot exceed 10\% of $|V_{us}|$. 
This result is largely independent from the value of $\phi_u$.

\begin{figure}[t]
\includegraphics[width=\columnwidth]{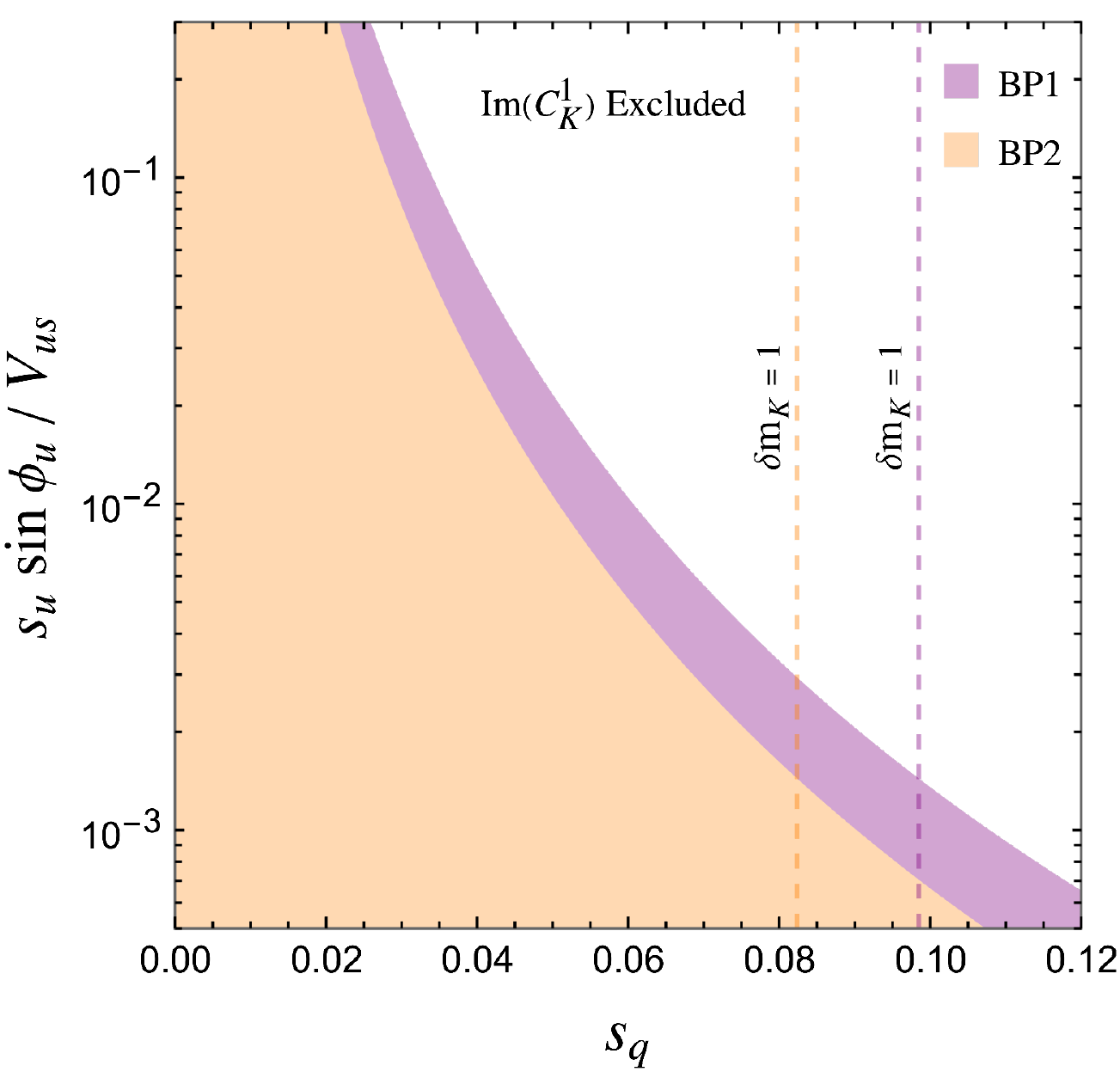}
\caption{Constraints from CP violation 
in  $K$--$\bar{K}$ mixing on the combination 
$s_u \sin\phi_u$ as a function of $s_q$.
The area outside the colored regions 
(corresponding to our two benchmark points) is excluded by the bounds on ${\rm Im} (C_K^1)$ at the $95\%$ confidence level. The vertical lines denote indicative upper bounds on $s_q$ set by  ${\rm Re}(C_K^1)$.
\label{fig:KKbarplot} }
\label{fig:branes}
\end{figure}

Assuming this mild up-alignment, we now move to the analysis of $K$--$\bar{K}$ mixing. Neglecting terms $O(s_{u}^2)$, the following simplified expression holds
 \begin{eqnarray}
{\rm Im}(C_K^1) &\approx&    \frac{8G_U}{\sqrt{2}}   \rho_q^{\rm NLO}   V_{ud}^3 V_{us} s_q^4 s_u \sin \phi_u \,, 
 \end{eqnarray} 
 and we see that the size of CP violation in kaon mixing is controlled by $s_q$ and the product $s_u \sin \phi_u$. 
As shown in \cref{fig:KKbarplot}, the latter 
product is severely constrained  for $|s_q| \gtrsim |V_{cb}|$.
However, it is worth noticing that 
$s_u \sin \phi_u = 0$
 is achieved either assuming $\phi_u=0$ 
(hence CP conservation)
or in the limit of exact up-alignment (i.e.~for $s_u=0$).


In principle, an alternative solution also exists, as shown in the upper half of \cref{fig:DDbarplot}. 
However, for this solution to pass both $K$--$\bar{K}$ and $D$--$\bar D$ mixing bounds, a 
 strong but not exact down-alignment is needed, 
  $|s_d| \approx {\rm Im}(V_{ub}^* V_{cb})/s_q^2$,
  and the phase difference $\Delta_{ud}$ 
(defined in \cref{app:appendixA}) need to be set to $\Delta_{ud} \sim {\rm sgn} (s_d\cos \phi_u)\,\pi/2$. We find 
these conditions more difficult to justify from a UV model point of view, being far from a clear symmetric limit.
For this reason, in the following we focus on the up-aligned solution.

In summary, to pass the bounds on ${\rm Im}(C_{K,D}^1)$ we require:
  i)~$|s_u| \ll  |V_{us}|$ and (depending on the degree of up-alignment) ii)~$\phi_u \approx 0$, corresponding to the assumptions of i)~up-alignment for the $2\times 2$ light-quark Yukawa couplings and ii)~approximate CP conservation.

Turning our attention now to the real parts of the $\Delta F=2$ amplitudes, in the limit  $s_u \approx 0$ 
only $K$--$\bar K$ mixing receives a sizable contribution
 \begin{equation}
 {\rm Re}(C_K^1) \approx \frac{4G_U}{\sqrt{2}}   \rho_q^{\rm NLO}    V_{ud}^2 V_{us}^2 s_q^4   \,,  \\
 \end{equation}
 which is dominantly controlled by $s_q$.
 This induces a non-standard  contribution to the $K$--$\bar K$ mass difference, that we can write as
 \begin{eqnarray}
\delta m_K = \frac{\Delta m^{\rm NP}_K}{\Delta m^{\rm exp}_K}  &=&  \frac{2}{3}\frac{ m_K}{ \Delta m^{\rm exp}_K } F_K^2   P_{1}^{\rm VLL}\,  {\rm Re}(C_K^1)  \,,
 \end{eqnarray} 
 where $m_K = 497.6$ MeV, $\Delta m^{\rm exp}_K = 3.484\times 10^{-12}$ MeV, $F_K = 160$ MeV, and $P_{1}^{\rm VLL}(\mu_K) = 0.48$ encodes the bag parameter and QCD corrections~\cite{Buras:2001ra}. Numerically, we find
  \begin{eqnarray}
 \delta m_K \approx 1.086 \times \left(  \frac{C_U }{0.015 } \right) \left( \frac{ s_q }{ 0.1} \right)^4\, .
 \end{eqnarray} 
 In view of the long-distance contributions to  $\Delta m_K$, we can translate this result into a bound on the size of $|s_q | \lsim 0.1$ for the preferred range of $C_U$. 
 
 \subsection{Summary on the mixing structure}
What emerges from the previous discussion can be summarized as follows:
\begin{itemize}
\item{\em Structure of the Yukawa couplings.} The mixing of light families in the CKM (i.e.~the Cabibbo angle) is induced mainly by the down Yukawa coupling ${\bf Y}_d$.
On the other hand, heavy $\leftrightarrow$ light mixing and the CKM phase are  generated mainly by ${\bf Y}_u$.
\item{\em Light $\leftrightarrow$ vector-like mixing.} The parameter $s_q$, 
which controls the largest breaking of $U(2)_q$ and plays a key role in all the LFU ratios, cannot exceed 0.1 in magnitude. 
\end{itemize}

\section{$K\rightarrow \pi \nu\bar\nu$}\label{sec:Kpinunubar}
We now proceed to analyze the implications of the mixing structure determined in \cref{sec:deltaF2} on $K\rightarrow \pi \nu\bar\nu$ decays.
In the EW-symmetric phase, we define the semi-leptonic operator
\begin{align}
\mathcal{O}_{\ell q}^{\alpha\beta ij}&=(\bar \ell_L^\alpha\gamma_\mu \ell_L^\beta)(\bar q_L^i\gamma^\mu q_L^j)\,.
\end{align}
Our setup is such that only $s\to d\nu\nu$ transitions involving third-generation neutrinos receive large NP corrections.\footnote{The leading correction is suppressed relative to the third-generation contribution by $s_\ell^2$. } Therefore, the relevant part of the effective Lagrangian reads
\begin{align}\label{eq:sdnn_SMEFT}
    \mathcal{L}_{\rm eff}  \supset -\frac{4G_U}{\sqrt{2}} \sum_{ij} \mathcal{C}_{\ell q}^{33ij}  \mathcal{O}_{\ell q}^{33ij}+{\rm h.c.}\,,
\end{align}
where the sum runs over $i,j=1,2,3$. 
The relevant contributions to this process come from the tree-level $Z'$ exchange for the flavor-conserving Wilson coefficients $\mathcal{C}_{\ell q}^{33ii}$ as well as from one-loop diagrams involving the $U_1$ leptoquark for flavor-conserving and flavor-violating Wilson coefficients $\mathcal{C}_{\ell q}^{33ij}$, with $i\land j = 2 \vee 3$. These contributions are reported in Table~\ref{tab:SLWCs} in~\cref{app:appendixB}.

Similarly to the case of $b\to s\nu\nu$ transitions (see e.g.~\cite{Barbieri:2015yvd,Fuentes-Martin:2020hvc})
there is no tree-level contribution mediated by the
$U_1$ LQ to $s\to d\nu\nu$ amplitudes. This is because the 
tree-level $U_1$ exchange 
generates both $SU(2)_L$-singlet semi-leptonic operators as in 
\eqref{eq:sdnn_SMEFT} as well as their $SU(2)_L$-triplet counterparts, $(\bar \ell_L^3\gamma_\mu \tau^I \ell_L^3)(\bar q_L^i\gamma^\mu \tau^I q_L^j)$, with the same Wilson coefficient, resulting in a cancellation. This structure is simply a consequence of the $U_1$ gauge quantum numbers, which forbid its coupling to down-type quarks and neutrinos.

After EW symmetry is broken, we project the contributions of \eqref{eq:sdnn_SMEFT} onto the coefficients of the operators $(\bar s_L \gamma_\mu d_L) (\bar\nu_{\ell}\gamma^\mu \nu_{\ell})$, that we normalize as in the SM
\begin{equation}
\mathcal{L}_{\rm eff} = -\frac{4 G_F}{\sqrt{2}}\frac{\alpha_W}{2\pi}\, V_{ts}^{*}V_{td}\, \sum_{\ell=e,\mu,\tau} C_{\ell}\, 
(\bar s_L \gamma_\mu d_L) (\bar\nu_{\ell}\gamma^\mu \nu_{\ell})\,,
\label{eq:sdnn_LEET}
\end{equation}
where $\alpha_W = g_2^2/(4\pi)$ and $V_{ij}$ are CKM matrix elements. 
Using the results in~\cite{Buras:2015qea}, 
we decompose $\cB(K^+ \to \pi^+ \nu \nu)$
and  $\cB(K_L \to \pi^0 \nu \nu)$
in terms of the $C_{\ell}$ as 
\begin{eqnarray}
&& \cB(K^+ \to \pi^+ \nu \nu) = \cB_+ 
 \sum_{\ell=e,\mu,\tau} 
\Big\{ \big[{\rm Im}(V_{ts}^* V_{td} C_\ell  )\big]^2 + \no \\ 
&& \hspace{1.65cm}
\left[
|V_{us}|^5 |V_{cs}| P_c -
{\rm Re}\left(V_{ts}^* V_{td} C_\ell \right) \right]^2   \Big\}~, \label{eq:BRKpnn}
\\
&& \cB(K_L \to \pi^0 \nu \nu) = \cB_L 
 \sum_{\ell=e,\mu,\tau} 
 \big[{\rm Im}(V_{ts}^* V_{td} C_\ell  )\big]^2~, 
\label{eq:BRKLpnn}
\end{eqnarray}
where $B_+ =2.62\times 10^{-6}/|V_{us}|^2$
and $B_L =4.16\times 10^{-6}/|V_{us}|^2$.
Within the SM, 
\begin{equation}
C^{{\rm SM}}_{e,\mu,\tau} \equiv X_t\,,
\label{eq:sdnn_CSM}
\end{equation}
with $X_t = 1.48\pm 0.01$~\cite{Buchalla:1998ba},
while the parameter encoding the long-distance contribution due to charm and light-quark loops (averaged over the different lepton species)
 is $P_c =0.404 \pm 0.024$.\footnote{In principle, 
 $P_c$ is different for the different lepton species;
 however, we have explicitly checked that this 
 effect is negligible compared to the 
 model  uncertainties affecting $\Delta C_{\tau}$.}

\begin{figure}[t]
\includegraphics[scale=1.15]{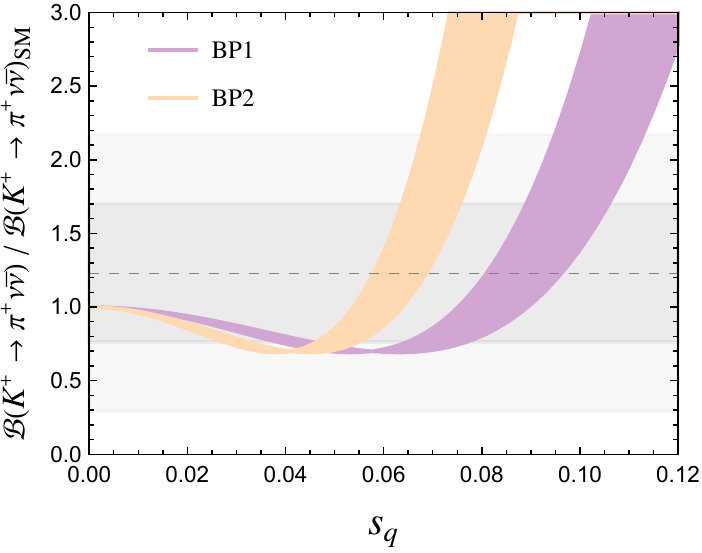}
\caption{
Prediction for $\cB(K^+ \rightarrow \pi^+ \nu\bar\nu )$, normalized to the SM, as a function of $s_q$ for the two benchmark points we have considered (setting  $\theta_\chi=\pi/4$, $m_L = 1$ TeV, $m_Q = 1.5$ TeV).
The colored bands are generated by varying $g_4 
\in [2,3]$ and $\xZ \in [0.8,1]$. 
The gray regions show the current experimental determination of $\cB(K^+ \rightarrow \pi^+ \nu\bar\nu )$ at $\pm 1\sigma$ and $\pm 2\sigma$.  }
\label{fig:Kpnn}
\end{figure}

Within our framework, (\ref{eq:BRKpnn})--(\ref{eq:BRKLpnn}) remain valid and the only modification compared to the SM case is the value of 
$C_{\tau}$, that we decompose as
\begin{equation}
C_{\tau}  \approx  C^{{\rm SM}}_{\tau}  \left[  1  +  \rho \,    \Delta C_{\tau} \right]\,.
\end{equation}
Running (\ref{eq:sdnn_SMEFT}) down to the EW scale, 
and working in the basis where the down Yukawa coupling is diagonal we get

\begin{align}\label{eq:DeltaCtau}
\Delta C_{\tau} &= \frac{1}{V_{ts}^{*}V_{td}} \sum_{ij} L_d^{si} \, \left[\mathcal{C}_{\ell q}^{33ij}+\beta_{L}^{i3} (\beta_{L}^{j3})^{*} \mathcal{C}_U^{\rm RGE} \right] (L_d^{dj})^{*} \,,\nonumber \\
\rho &= \frac{2\pi}{\alpha_W X_t} C_U  = 1.26 \times \left(\frac{C_U}{0.01}\right)\,.
\end{align}
The contributions of the flavor-violating Wilson coefficients to~\eqref{eq:DeltaCtau} are suppressed by the 23 and 13 down rotations, which are required to be small in order to evade the bounds on $B_{s(d)}$ mixing.
As a consequence, the sum over $\mathcal{C}_{\ell q}^{33ij}$ is dominated by the flavor conserving Wilson coefficients, which receive a Cabbibo-sized rotation
\begin{align}
 \sum_{ij} L_d^{si} \, \mathcal{C}_{\ell q}^{33ij} (L_d^{dj})^{*} \approx (\mathcal{C}_{\ell q}^{3311}-\mathcal{C}_{\ell q}^{3322})  \, c_d s_d e^{-i\phi_d} \,.
\end{align}
The leading correction to this result
that, as expected,  vanishes in the $U(2)_q$ symmetric limit, 
is $\mathcal{O}(s_b/s_q)$.
\begin{figure*}[t]
\includegraphics[scale=1.15]{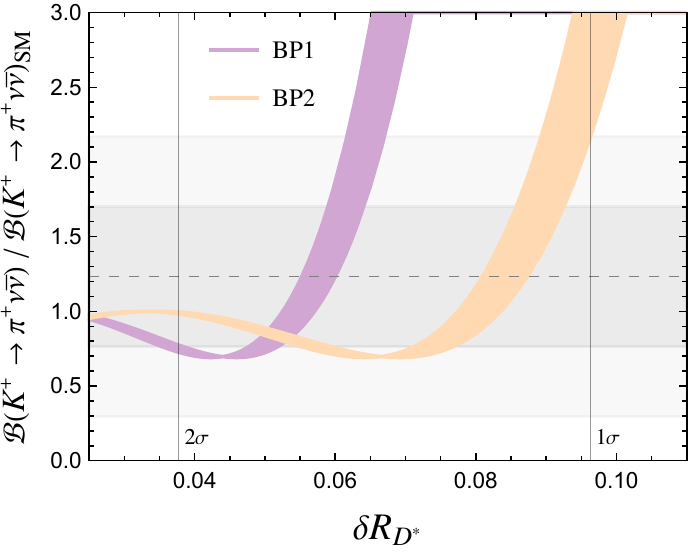}
 \hspace{10mm}
\includegraphics[scale=1.15]{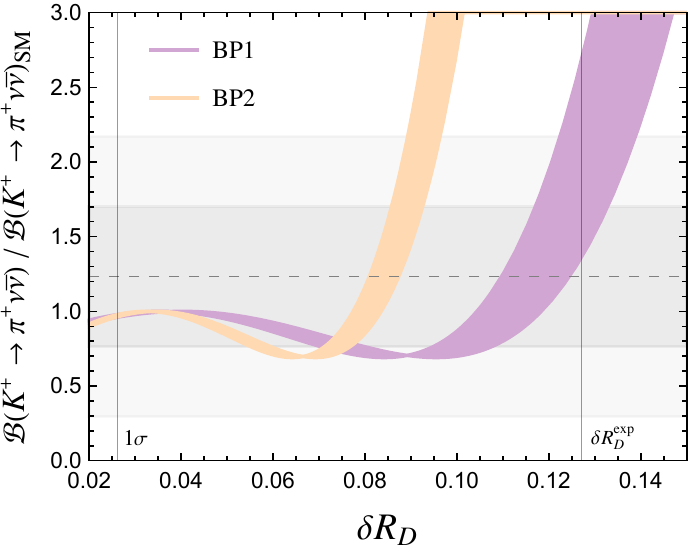}
\caption{
Prediction for $\cB(K^+ \rightarrow \pi^+ \nu\bar\nu )$, normalized to the SM, vs.~the relative modifications of the LFU ratios $R_{D^*}$ (left) and $R_{D}$ (right), also normalized to the SM. The notation and parameters are the same as in 
\cref{fig:Kpnn}, and the colored bands are again generated by varying $g_4 
\in [2,3]$ and $\xZ \in [0.8,1]$. The vertical lines indicate the 
current experimental determination of the LFU ratios
at different levels of significance.}
\label{fig:KpnnRd}
\end{figure*}
Similarly, the leading contribution to the $\mathcal{C}_U^{\rm RGE} $ term encoding the RG evolution of tree-level leptoquark mediated operators is given by the $i=j=2$ term 
\begin{equation}
L_d^{s2}\beta_{L}^{23}(\beta_{L}^{23}L_{d}^{d2})^{*} = -c_d s_d  e^{-i\phi_d} |W_{21}|^2 s_q^2 \,.
\end{equation}
Using DsixTools~\cite{Celis:2017hod} and running down from $\mU=2.5~\mathrm{TeV}$, we find
\begin{align}
\mathcal{C}_U^{\rm RGE}\approx-0.05\,.
\end{align}
Using the CKM relation $V_{us}  = e^{i\phi_d} c_u s_d - e^{i\phi_u} c_d s_u$ from~\cref{eq:CKMelems}, we obtain a final expression for 
$\Delta C_{\tau}$ that is dominantly controlled by $s_q$ in the up-aligned limit:
\begin{align}
\Delta C_{\tau} &\approx s_q^2 \frac{V_{us}^{*} V_{ud} }{V_{ts}^{*} V_{td}} \left[ \frac{1}{4x_{Z'}} + \frac{\alpha_4}{4\pi} B_{q\ell}^{2211} -|W_{21}|^2 \mathcal{C}_U^{\rm RGE} \right] \,. 
\end{align}
Importantly, the phase of $\Delta C_{\tau}$ is fully specified, leading to the following conclusions:
\begin{itemize}
\item ${\rm Im}(V_{ts}^* V_{td} \Delta C_{\tau} )\approx 0\ \rightarrow\ $
{\em negligible correction} compared to the SM in 
$\cB(K_L \rightarrow \pi^0 \nu\bar\nu )$;
\item  ${\rm Re}(V_{ts}^* V_{td} \Delta C_{\tau} ) > 0\ \rightarrow\ $
 {\em negative interference} between NP and SM amplitudes in
$\cB(K^+ \rightarrow \pi^+ \nu\bar\nu )$. 
\end{itemize}

The prediction of $\cB(K^+ \rightarrow \pi^+ \nu\bar\nu )$ thus obtained, as a function of $s_q$, for the two benchmarks defined in Section~\ref{sec:deltaF2}
is shown in   \cref{fig:Kpnn}. As can be seen,   $\cB(K^+ \rightarrow \pi^+ \nu\bar\nu )$ exhibits a strong and well-defined dependence 
 on $s_q$, which will be extremely valuable in determining this parameter from data in the near future.  
 Interestingly enough, the current experimental measurement of $\cB(K^+ \rightarrow \pi^+ \nu\bar\nu )$~\cite{NA62:2021zjw}, even if
affected by a sizable  error, 
already provides a non-trivial constraint on the model. In fact, for benchmark point 2, the bound on $|s_q|$
we determine from  $\cB(K^+ \rightarrow \pi^+ \nu\bar\nu )$ is more stringent than the one derived from $\Delta F=2$ observables. 

In \cref{fig:KpnnRd}, we show the correlation between 
 $\cB(K^+ \rightarrow \pi^+ \nu\bar\nu )$ and the two LFU ratios in charged-current $B$ decays, which also exhibit a strong (but different) dependence on $s_q$. In particular, what we show in \cref{fig:KpnnRd} are the relative corrections 
 \begin{equation}
    \delta R_{D^{(*)}} =  \frac{ R_{D^{(*)}} }{ R_{D^{(*)}}^{\rm SM}  } -1~,
 \end{equation}
 that we compute using the NLO expressions~\cite{Fuentes-Martin:2020hvc} 
 reported in Appendix~\ref{app:appendixD}. Contrary to all the observables
 discussed so far, $R_{D}$ and $R_{D^{*}}$ are sensitive to the 
 coupling of the leptoquark field to third-generation RH chiral fermions
 (parameterized in terms of the effective coupling $\beta_R^{b\tau}$).
In the minimal version of the model, $|\beta_R^{b\tau}|=1$; 
however, in less minimal frameworks
different values could be obtained with
additional sets of VL fermions.\footnote{One example is the introduction of $SU(3)_l$-charged VL partners of the light family right-handed fields, as in \cref{app:appendixC}.} 
Moreover, in general $\beta_{R}^{b\tau}$
has a non-trival phase, $\arg (\beta_{R}^{b\tau})=-\arg (y_\tau/y_b)$, that appears after re-phasing $b_R$ and $\tau_R$ to make the corresponding Yukawa couplings $y_{b,\tau}$ real.
 For illustrative purposes and also to connect with previous studies~\cite{Cornella:2021sby}, 
 in the following we consider two choices for $\beta_R^{b\tau}$
 in relation to the two benchmarks considered so far:
 \begin{align}
&{\rm BP1}~[C_U = 0.015]: \qquad  \beta_R^{b\tau} =-0.3~, \\
&{\rm BP2}~[C_U = 0.030]: \qquad  \beta_R^{b\tau} =0~.
\end{align}
As can be seen in Appendix~\ref{app:appendixD}, a non-vanishing ${\rm Re}(\beta_R^{b\tau})$ mainly impacts $R_{D}$, which is particularly sensitive to scalar-current contributions.

This is illustrated by the comparison 
of left and right plots in \cref{fig:KpnnRd}. We finally stress that the value of 
$C_U$, which is the main parameter distinguishing the different curves in 
\cref{fig:KpnnRd}, could in principle be determined from high-energy data, 
in particular from $\sigma(pp \to \tau^+\tau^-)$~\cite{Cornella:2021sby}. 
In the future, the combination of precision measurements of 
$R_{D}$, $R_{D^{*}}$, and  $\cB(K^+ \rightarrow \pi^+ \nu\bar\nu )$,
combined with high-energy data on $\sigma(pp \to \tau^+\tau^-)$,
should result in decisive tests of this framework.

\section{Conclusions}\label{sec:conclusions}
  
 The  non-universal 4321 gauge model featuring third-family quark-lepton unification represents a very attractive extension of the SM. 
On the theoretical side, it can arise as the low-energy limit of ambitious UV models
addressing both the origin of quark and lepton mass hierarchies, as well as the electroweak hierarchy problem~\cite{Bordone:2017bld,Fuentes-Martin:2020bnh,Fuentes-Martin:2022xnb}. 
On the phenomenological side, it predicts the existence of a TeV-scale vector leptoquark that could account for the hints of LFU violation observed in semi-leptonic $B$-meson decays. 

As we have shown, the minimal version of the model is quite constrained by low-energy observables, resulting in a predictive framework not only for the leading $3 \to 2$ 
flavor-changing transitions, but also for processes involving light quarks only. 
In this context, precise measurements of $K \to \pi \nu \bar\nu$ decay  rates can 
play a key role in  further testing this setup.

Confirming previous findings obtained in different SM extensions~\cite{Bordone:2017lsy,Descotes-Genon:2020buf,Fajfer:2018bfj,Marzocca:2021miv}, 
we find sizable deviations from the SM in $\cB(K^+ \to \pi^+ \nu \nu)$, within the parameter space motivated by the $B$-meson anomalies.
Most importantly, as illustrated in \cref{fig:KpnnRd}, a well-defined pattern of 
correlations between $\cB(K^+ \to \pi^+ \nu \nu)$, $R_{D}$, and $R_{D^{*}}$, emerges. 
Testing these correlations, with future precision measurements of these 
low-energy observables, 
would provide an extremely valuable test of this motivated framework.

\section*{Acknowledgements}
This project has received funding from the European Research Council (ERC) under the European Union's Horizon 2020 research and innovation programme under grant agreement 833280 (FLAY), and by the Swiss National Science Foundation (SNF) under contract 200020\_204428.

\appendix

\section{4321 Gauge Boson Couplings}\label{app:appendixGB}
Referring to the currents in Eqs.~\ref{eq:ULag}-\ref{eq:ULagEnd} and defining $\eps_{3} = g_3^2/g_4^2$, $\eps_{1} = 2g_1^2/3g_4^2$, the 4321 gauge boson couplings in the VL fermion mass basis are given as
\begin{align}
\kappa_q &= {\bf O}_{q} \kappa_q^\prime {\bf O}_{q}^{\dagger} \,, \hspace{8mm} \kappa_q^\prime=\mathrm{diag}(-\eps_3,-\eps_3,1,1)\,, \\
\xi_q &= {\bf O}_{q} \, \xi_{q}^\prime {\bf O}_{q}^{\dagger} \,, \,  \hspace{8mm} \xi_q^\prime=\mathrm{diag}(-\eps_1,-\eps_1,1,1)\,, \\
\xi_\ell &= {\bf O}_{\ell} \, \xi_{\ell}^\prime \, {\bf O}_{\ell}^{\dagger} \,, \, \hspace{8mm} \xi_\ell^\prime=\mathrm{diag}(-\eps_1,-\eps_1,1,1)\,,
\end{align}
and
\begin{equation}
\beta_L = {\bf O}_{q} {\bf W} \beta_L^\prime {\bf O}_{\ell}^{\dagger} \,, \hspace{8mm} 
{\bf W} = 
\begin{pmatrix}
\mathbb{1}_{2\times 2} & 0_{2\times 2}  \\
0_{2\times 2} & W  \\
\end{pmatrix}   \,,
\end{equation}
with $\beta_L^\prime=\mathrm{diag}(0,0,1,1)$. For the neutral gauge bosons $G', Z'$, we find explicitly
\begin{align}
\kappa_q &=
\left(\begin{array}{ccc|c}
-\eps_{3} & 0 & 0 & 0 \\
0 &  - \eps_{3} + s_{q}^2  & 0 &  -c_{q} s_{q}  \\
0 & 0 & 1 &  0 \\ [2pt] \hline 
0& -c_{q} s_{q} & 0 & c_{q}^2  \\
\end{array}\right) 
\,, \\
\xi_{q,\ell}  &=
\left(\begin{array}{ccc|c}
-\eps_1 & 0 & 0 & 0 \\
0 & -\eps_1 + s_{q,\ell}^{2}  & 0 &  -c_{q,\ell} s_{q,\ell}  \\
0 & 0 & 1 &  0 \\ [2pt] \hline 
0& - c_{q,\ell} s_{q,\ell}  & 0 & c_{q,\ell}^2  \\
\end{array}\right) 
\,,
\end{align}
where we have dropped subleading terms $\propto \eps_{1,3}\,  s_{\ell,q}\, $. Similarly, for the charged (LQ) current, we find
\begin{align}
\begin{aligned}
\beta_{L} = 
\left(\begin{array}{ccc|c}
0 & 0 & 0 & 0 \\
0 & c_{\chi} s_{q} s_{\ell} & s_{\chi} s_{q} &  -c_{\chi} s_{q} c_{\ell}  \\
0 & - s_{\chi} s_{\ell} & c_{\chi} &  s_{\chi} c_\ell \\ [2pt] \hline 
0& - c_{\chi} c_{q} s_{\ell}& - s_{\chi} c_q & c_{\chi} c_{q} c_{\ell} \\
\end{array}\right) 
\,.
\end{aligned}
\end{align}
From these results, we can also easily obtain the couplings of the 4321 gauge bosons in the basis where the SM down quark and charged lepton Yukawas are diagonal
\begin{align}
\kappa_q^{\rm down} &= L_d \, \kappa_q \, L_d^{\dagger} \,, \\
\xi_{q,\ell}^{\rm down} &= L_{d,e} \, \xi_{q,\ell} \, L_{d,e}^{\dagger} \,, \\
\beta_{L}^{\rm down} &= L_d \, \beta_{L} \, L_e^{\dagger} \,,
\end{align}
where the left-handed rotation matrices $L_{d,e}$ are defined in Appendix~\ref{app:appendixA}.

\section{Yukawa Diagonalization and the CKM Matrix}\label{app:appendixA}
The Yukawas can be brought to diagonal form in the $3\times3$ sub-block by rotations of the form $L_u  {\bf Y}_u R_u^{\dagger} = \hat {\bf Y}_u$, $L_d  {\bf Y}_d R_d^{\dagger} = \hat {\bf Y}_d$, and $L_e  {\bf Y}_e R_e^{\dagger} = \hat {\bf Y}_e$. In terms of these rotation matrices, the Lagrangian takes the form
\begin{align}
\begin{aligned}
-\mathcal{L} & \supset \bar \Psi_{q} L_u^{\dagger} \hat {\bf Y}_u R_u \tilde{H}  \Psi_{u} +  \bar \Psi_{q}  L_d^{\dagger} \hat{\bf Y}_d R_d H \Psi_{d}  \\
&+ \bar \Psi_{\ell} L_e^{\dagger} \hat{\bf Y}_e R_e H \Psi_{e}  + \textrm{h.c.} \,.
\end{aligned}
\end{align}
These rotation matrices can be decomposed as the product of rotations in the 
$13 \times 23 \times 12$ sectors. 
For example, in the up sector we have
\begin{equation}
L_u = P_u L_{u}^{12} L_{u}^{23} L_u^{13} \,, 
\end{equation}

with
\begin{align}
\begin{aligned}
L_u^{12} &= 
\left(\begin{array}{cccc}
c_u & -e^{i \varphi_u} s_u & 0 & 0 \\
e^{-i \varphi_u} s_u & c_u   & 0 & 0 \\
0 & 0 & 1 & 0\\ 
0 & 0 & 0 & 1\\
\end{array}\right) 
\,, \\
L_{u}^{13} &=
\left(\begin{array}{cccc}
c_{ut} & 0& -e^{i \varphi_{ut}} s_{ut} & 0 \\
0  &  1   & 0 & 0 \\
e^{-i \varphi_{ut}} s_{ut} & 0 & c_{ut} & 0\\ 
0 & 0 & 0 & 1\\
\end{array}\right)\,, \\
L_{u}^{23} &=
\left(\begin{array}{cccc}
1 & 0 & 0 & 0 \\
0 & c_t   & -e^{i\varphi_t} s_t & 0 \\
0 & e^{-i\varphi_t} s_t & c_t  & 0\\ 
0 & 0 & 0 & 1\\
\end{array}\right) \,,
\end{aligned}
\end{align}
and $P_{u,d}$ are arbitrary phase diagonal matrices that we will fix later to match the standard CKM phase convention.
The matching of the Lagrangian parameters in (\ref{eq:Yu}) onto the rotation matrix angles in 
$L_{u}^{23}$ and $L_{u}^{13}$ is 
\begin{equation}
s_t = -x_t s_q~, \qquad e^{i \varphi_{ut}}  s_{ut}=- e^{i \varphi_{U}} \eps_U~,
\end{equation}
while $L_{u}^{12}$ is in direct correspondence with the $2\times 2$ matrix $U_u$.

There are also chirally suppressed RH rotations
\begin{align}
\begin{aligned}
R_{u}^{23} &=
\left(\begin{array}{cccc}
1 & 0 & 0  \\
0 & 1  & -\frac{m_c}{m_t} e^{i\varphi_t} s_t  \\
0 & \frac{m_c}{m_t}e^{-i\varphi_t} s_t & 1 \\ 
\end{array}\right)\,, \\
R_{u}^{13} &=
\left(\begin{array}{cccc}
1 & 0&  -\frac{m_u}{m_t}e^{i \varphi_{ut}} s_{ut}  \\
0  &  1   & 0  \\
\frac{m_u}{m_t}e^{-i \varphi_{ut}} s_{ut} & 0 & 1 \\ 
\end{array}\right)\,,
\end{aligned}
\end{align}
and $R_u=P_u R^{23}_u R^{13}_u$.
The down quark and charged lepton rotations follow the same pattern, and can easily be written in terms of $s_d$, $s_b$, $s_{db}$ and $s_e$, $s_\tau$, $s_{e\tau}$ with their corresponding phases.

\subsection{CKM Matrix}

To compute the CKM matrix,
\begin{equation}
V_{\rm CKM} = L_u L_d^{\dagger} \,,
\end{equation}
we will work with the rotation matrices defined in the previous section and keep the angles arbitrary. The CKM computed in this way will be valid for any version of the model, since all that changes will be the matching of the model parameters onto the angles.
We will compute the CKM matrix to $\mathcal{O}(\lambda^3)$ assuming the following power counting for the rotation matrix angles
\begin{align}
\theta_{u,d} \sim \lambda  \,,\,\, \theta_{t} \sim \lambda^2 \,,\,\,  \theta_b \sim \lambda^3\,,\,\,  \theta_{ut} \sim \lambda^3\,,\,\,  \theta_{db} \sim \lambda^4 \,,\label{eq:PowerCountAng}
\end{align}
where $\lambda=|V_{us}|=0.225$.
It is convenient to define the following phases
\begin{align}
\Delta_{ud} =&\, \varphi_u - \varphi_d\,, \\
\Delta_{tb} =&\, \varphi_t - \varphi_b\,, \\
\Delta_{ut} =&\,\varphi_{ut}-\varphi_{u}-\varphi_t\,,\\
\phi_1=& \arg( c_u c_d + e^{-i\Delta_{ud}}s_u s_d )\,,\\
\phi_u =& \arg (e^{i\Delta_{ud}}c_u s_d -s_uc_d)\,, \\ 
\phi_t =& \arg (-s_t+e^{i\Delta_{tb}}s_b) \,, \\
\phi_{ut} =& \arg (s_u s_t -e^{i\Delta_{ut}}s_{ut} ) \,, \\
\phi_d =& \phi_u-\Delta_{ud}\,, \\
\phi_b =&\phi_t-\Delta_{tb}\,.
\end{align}
We see that according to our power counting, if $\Delta_{ud}$, $\Delta_{tb}$ and $\Delta_{ut}$ are order one, then we have  $\phi_1 \sim \mathcal{O}(\lambda^2)$, $ \sin \phi_t \sim \mathcal{O}(\lambda)$, and $  \phi_{ut},\, \phi_{u}, \phi_d, \phi_b  \sim \mathcal{O}(1)$. 

Choosing the arbitrary phase matrices 
\begin{align}
P_u &= {\rm diag} \{ e^{i(\phi_u-\varphi_u+\phi_t-\varphi_t+ \phi_1)},\,e^{i (\phi_t-\varphi_t)},\, 1,\, 1 \}\,, \\
P_d &= {\rm diag} \{ e^{i(\phi_u-\varphi_u+\phi_t-\varphi_t)},\,e^{i (\phi_t-\varphi_t+\phi_1)},\, 1,\, 1 \} \,,\label{eq:Pud}
\end{align}
we find that the CKM takes the following form 
\begin{align}
\begin{aligned}
V_{\rm CKM} =
\left(\begin{array}{ccc|c}
V_{ud} &  V_{us} &   V_{ub} & 0  \\
V_{cd}    & V_{cs}    & V_{cb} & 0 \\ 
V_{td}& V_{ts} & V_{tb} & 0 \\ \hline 
0 & 0 & 0 & 1 \\ 
\end{array}\right)  \,, 
\end{aligned}
\end{align}
where, at $\mathcal{O}(\lambda^3)$, we have
\begin{align}
\begin{aligned}
V_{ud}&= e^{i\phi_1} (c_u c_d + e^{i\Delta_{ud}}s_u s_d )\\
V_{us}&=e^{i \phi_u} (e^{-i \Delta_{ud}} c_u s_d - s_u c_d ) \\
V_{ub}&=e^{i (\phi_u+\phi_t+\phi_1)}(s_u s_t - e^{i\Delta_{ut}} s_{ut}  )\\
V_{cd}&= -V_{us}^* \\
V_{cs}&= V_{ud}^* \\
V_{cb}&=e^{i\phi_t} (-s_t + e^{-i \Delta_{tb}} s_b) \\
V_{td} &= e^{-i(\phi_u+\phi_t)}(-e^{i\Delta_{ud}}s_ds_t  + e^{-i\Delta_{ut}} s_{ut}) \\
V_{ts}&= -e^{-i\phi_1} V_{cb}^* \\
V_{tb} &= 1\,.
\end{aligned}
\label{eq:CKMelems}
\end{align}
We have chosen the phases in~\eqref{eq:Pud} such that $V_{ud}$, $V_{us}$, and $V_{cb}$ are real.
At this order in $\lambda$, we have 
$V_{td}\approx V_{us}^{*} V_{cb}^{*}-V^{*}_{ub}$.
The CKM phase, $V_{ub}=|V_{ub}|e^{-\delta}$, is then
\begin{align}
\delta=-( \phi_{ut}+ \phi_u+\phi_t+ \phi_1).
\end{align}

\subsection{Quark Rotation Matrices}
Neglecting the small $\phi_1$ phase
and working to $\mathcal{O}(\lambda^3)$, we find the quark rotation matrices to be

\vspace{0.5cm}

\noindent \hspace{1mm} $L_d = P_d L_{d}^{12} L_{d}^{23} L_d^{13}=$
\begin{align}
\begin{aligned}
\left(\begin{array}{ccc|c}
c_d & -e^{i \phi_{d}} s_d  & 0 & 0  \\
e^{-i \phi_{d}} s_d  &  c_d & -e^{i\phi_b} s_b  & 0 \\
0 &  e^{-i \phi_b} s_b & 1 &  0 \\ \hline
0 & 0 & 0  & 1 \\
\end{array}\right) P_q
\,,
\end{aligned}
\end{align}
\hspace{1mm}$L_u =P_u L_u^{12} L_u^{23} L_u^{13}=$
\begin{align}
\begin{aligned}
 \left(\begin{array}{ccc|c}
c_u & -e^{i\phi_{u}} s_u  & |V_{ub}| e^{-i \delta} & 0  \\
e^{-i  \phi_{u}} s_u  & c_u  & -e^{i\phi_t} s_t  & 0 \\
 e^{-i( \phi_{u}+ \phi_{t})} s_{u}s_{t}-|V_{ub}|e^{i\delta}  &  e^{-i \phi_t} s_t & 1 &  0 \\ \hline
0 & 0 & 0  & 1
\end{array}\right) 
P_q
\,,
\end{aligned}
\end{align}
where
\begin{align}
P_q &= {\rm diag} \{ e^{i(\phi_u-\varphi_u+\phi_t-\varphi_t)},\,e^{i (\phi_t-\varphi_t)},\, 1,\, 1 \}\,.
\end{align}

\section{$SU(3)_l$ Vector-like Fermions}\label{app:appendixC}
We extend the matter content given in Table~\ref{tab:fieldcontent} by a new vector-like quark with quantum numbers $U_{L,R} \sim ({\bf 1, 3, 1}, 2/3)$ under $\mathcal{G}_{4321}$. The addition of this state extends the flavor symmetry as $U(2)_{u_R} \rightarrow U(3)_{\mathcal{U}_{R}} \times U(1)_{U_L} $, where we have defined a vector $\mathcal{U}_{R}^\intercal = (u_{R}^1 \; u_{R}^2  \; U_{R})$.  The new terms one can write in the Lagrangian are
\begin{align}\label{eq:newVLF}
\begin{aligned}
-\mathcal{L} & \supset  \bar U_L \hat{M}_{U} \, \mathcal{U}_{R}   +   \,  \lambda_{U} \bar U_{L} \Omega_{3} \psi_{R}^{+}  +  \bar q_{L}  {\bf y}_U  \tilde{H}    U_{R}  + \textrm{h.c.} \,,
\end{aligned}
\end{align}
where without loss of generality we can use the $U(3)_{\mathcal{U}_R}$ to put the $1\times 3$ mass  in the form $\hat{M}_{U} = (0 \; 0 \; m_U)$ with $m_U$ real and the $U(1)_{U_L}$ to make the complex number $\lambda_{U}$ real. In this basis, ${\bf y}_U$ is an arbitrary $2\times 1$ complex vector, where we define ${\bf y}_U^\intercal = (y_U^1 e^{i\varphi_U^1} \;  y_U^2 e^{i\varphi_U^2})$.  In total, we removed 4 phases and 2 real parameters, which is consistent since the flavor symmetry was enlarged from $U(2)_{u_R} \rightarrow U(3)_{\mathcal{U}_{R}} \times U(1)_{U_L}$, meaning we are allowed to remove $6+1-3 = 4$ phases and $3-1 = 2$ real parameters. Finally, it is clear that this part of the Lagrangian leaves a $U(2)_{u_R}$ subgroup unbroken, while the coupling ${\bf y}_U$ is a new source of $U(2)_q$ symmetry breaking.

\subsection{4321 Broken Phase}
After 4321 symmetry breaking, let us define $\Psi_{u}^\intercal = (u_{R}^1 \; u_{R}^2 \; u_R^3 \;  U_R)$. We then have to diagonalize the mixing between $u_R^3$ and $U_R$
\begin{equation}
\bar U_L
\begin{pmatrix}
0 & 0 & \frac{\lambda_{U} v_{3}}{\sqrt{2}} & m_U \\
\end{pmatrix} \Psi_{u} =
\bar U_L
\begin{pmatrix}
0 & 0 & 0 & m_U^{\rm phys} \\
\end{pmatrix} W_u \Psi_{u} \,,
\end{equation}
where
\begin{equation}
m_{U}^{\rm phys} = \sqrt{ m_{U}^2 + \frac{1}{2}\lambda_{U}^2 v_{3}^2} \,, 
\end{equation}
and $W_{u}$ is an orthogonal rotation mixing $u_R^3$ and $U_R$
\begin{equation}
W_{u} =  \begin{pmatrix}
1 & 0 & 0  & 0\\
0 & 1 & 0 & 0 \\
0& 0 & c_{U} & -s_{U} \\
0 & 0& s_{U} & c_{U} 
\end{pmatrix} \,,
\end{equation}
where the mixing angle is defined as
\begin{align}
\tan \theta_{U} &= \frac{\lambda_{U} v_{3}}{ \sqrt{2} m_{U}} \,.
\end{align}

\subsection{Yukawas}
After rotating to the VL fermion mass basis, the up Yukawa
\begin{align}
\begin{aligned}
-\mathcal{L} & \supset \bar \Psi_{q} {\bf Y}_u \tilde{H} \Psi_{u} + \textrm{h.c.} \,,
\end{aligned}
\end{align}
can be written as a $4\times 4$ matrix

\begin{widetext}

\begin{align}
\begin{aligned}
{\bf Y}_u &= {\bf O}_{q}
\left(\begin{array}{cc|cc}
U_{u}^{\dagger}   \hat Y_u  & 0 & {\bf y}_U \\
0  & y_t  & 0  \\  \hline 
0  & y_t x_{t}  e^{i \varphi_{t}}  & 0 \\
\end{array}\right) W_u^{\dagger} \\
&={\bf O}_{q} \left(\begin{array}{ccc|c}
y_u c_u & y_c  e^{i \varphi_u} s_u & -y_U^1 e^{i\varphi_{U}^1} s_U & y_U^1 e^{i\varphi_{U}^1} c_U  \\
-y_u  e^{-i \varphi_u} s_u  & y_c c_u    & -y_U^2 e^{i\varphi_{U}^2} s_U & y_U^2 e^{i\phi_{U}^2} c_U \\
0 & 0 & y_t c_U & y_t s_U\\ \hline 
0 & 0 & y_t x_{t}  e^{i \varphi_{t}} c_U  & y_t x_{t}  e^{i \varphi_{t}} s_U  \\
\end{array}\right) \\
&=\left(\begin{array}{ccc|c}
y_u c_u & y_c  e^{i \varphi_u} s_u & -y_U^1 e^{i\varphi_{U}^1} s_U & y_U^1 e^{i\varphi_{U}^1} c_U  \\
-y_u  e^{-i \varphi_u} s_u c_q & y_c c_u c_q   & -y_U^2 e^{i\varphi_{U}^2} s_U c_q -y_t x_{t}  e^{i \varphi_{t}} c_U s_q  & y_U^2 e^{i\varphi_{U}^2} c_U c_q -y_t x_{t}  e^{i \varphi_{t}} s_U s_q \\
0 & 0 & y_t c_U & y_t s_U\\  \hline 
-y_u e^{-i \varphi_{u}} s_u s_q & y_{c} c_u s_q & y_t x_{t}  e^{i \varphi_{t}} c_U c_q -y_U^2 e^{i\varphi_{U}^2} s_U s_q   & y_t x_{t}  e^{i \varphi_{t}} s_U c_q +y_U^2 e^{i\varphi_{U}^2} c_U s_q    \\
\end{array}\right) 
\,.
\end{aligned}
\label{eq:AppYu}
\end{align}
In the limit of small $\theta_U$ (corresponding to integrating out the new vector-like state $U_{L,R}$), the $4 \times 3$ sub-block reads
\begin{align}
\begin{aligned}
{\bf Y}_u 
&=\left(\begin{array}{ccc}
y_u c_u & y_c  e^{i \varphi_u} s_u & -y_U^1 e^{i\varphi_{U}^1} s_U  \\
-y_u  e^{-i \varphi_u} s_u c_q & y_c c_u c_q   & -y_{t} x_{t} e^{i \varphi_{t}}  s_q   \\
0 & 0 & y_t \\ \hline
-y_u e^{-i \varphi_{u}} s_u s_q & y_{c} c_u s_q & y_{t} x_{t} e^{i \varphi_{t}}  c_q    \\
\end{array}\right) 
\,,
\end{aligned}
\end{align}
which reproduces the structure we wanted in Section~\ref{sec:YukCouplings} and we identify $y_t \eps_U e^{i\varphi_U} = y_U^1 e^{i\varphi_{U}^1} s_U $. We therefore generate $V_{ub}$ with the right size for 
\begin{align}
y_{U}^1 \sin \theta_{U}  \approx  \frac{y_{U}^1 \lambda_{U} }{\sqrt{2}}\frac{v_{3}}{  m_{U}} \approx |V_{ub}| \,.
\end{align}

As a final comment, we note that in principle one could add a complete set of VL partners $U_{L,R}, N_{L,R}, D_{L,R}, E_{L,R}$ with the same quantum numbers as their corresponding light family RH fields. In this case, another interesting limit is that of large mixing (small cosine $c_{D,E}$) for the down-type partners, which could be exploited to explain the smallness of the down-type Yukawas while simultaneously reducing the magnitude of RH currents in the down-sector.

\section{Expressions for $R_{D^{(*)}}$}\label{app:appendixD}
Defining $\delta R_{D^{(*)}} = R_{D^{(*)}}/R^{\rm SM}_{D^{(*)}}-1$, we have
\begin{align}
\label{eq:dRD}\delta R_{D} &\approx 2C_U\, {\rm Re} \Big[ \rho_{LL}^{\rm NLO} \beta_{L}^{b\tau}\beta_{L}^{b\tau*} -1.50\,\eta_S\,\rho_{LR}^{\rm NLO}\,\beta_{L}^{b\tau}\beta_{R}^{b\tau*}\Big] \,,\\
\label{eq:dRDstar}\delta R_{D^*} &\approx 2C_U\, {\rm Re} \Big[ \rho_{LL}^{\rm NLO} \beta_{L}^{b\tau}\beta_{L}^{b\tau*} -0.12\,\eta_S\,\rho_{LR}^{\rm NLO}\,\beta_{L}^{b\tau}\beta_{R}^{b\tau*}\Big] \,,
\end{align}
where $\eta_{S} = 1.7$ and 
\begin{align}
\rho_{LL}^{\rm NLO} = \bigg(\rho_{LL,33}^{\rm NLO}+\rho_{LL,23}^{\rm NLO}\,A \bigg)~, \qquad 
\rho_{LR}^{\rm NLO} = \bigg(\rho_{LR,33}^{\rm NLO}+\rho_{LR,23}^{\rm NLO}\,A \bigg)~, \qquad 
    A = \bigg(\frac{V_{cs}}{V_{cb}} \frac{\beta_{L}^{s\tau}}{\beta_{L}^{b\tau}}+\frac{V_{cd}}{V_{cb}} \frac{\beta_{L}^{d\tau}}{\beta_{L}^{b\tau}}\bigg)\,.
\end{align}

The NLO corrections to $\mathcal{O}^U_{LL}$ and $\mathcal{O}^U_{LR}$ contributing to $b\to c\tau\nu$ transitions read \cite{Fuentes-Martin:2019ign}
\begin{align}
    \rho_{LL,33}^{\rm NLO} &= 1+\frac{\alpha_4}{4\pi}\left(\frac{17}{12}f_1(x_{Z^\prime})+\frac{4}{3}f_1(x_{G^\prime})\right)\,,\\
    \rho_{LL,23}^{\rm NLO} &= 1+\frac{\alpha_4}{4\pi}\bigg(\frac{7}{8}f_1(x_{Z^\prime})+\left(\frac{13}{24}\,f_2(x_{Z^\prime},x_Q)+\frac{4}{3}\,f_2(x_{G^\prime},x_Q)\right)c_q^2\,\bigg)\,,\\
    \rho_{LR,33}^{\rm NLO} &= 1+\frac{\alpha_4}{4\pi}\left(\frac{23}{12}f_1(x_{Z^\prime})+\frac{16}{3}f_1(x_{G^\prime})\right)\,,\\
    \rho_{LR,23}^{\rm NLO} &= 1+\frac{\alpha_4}{4\pi}\bigg(\frac{13}{8}f_1(x_{Z^\prime})+\left(\frac{7}{24}\,f_2(x_{Z^\prime},x_Q)+\frac{16}{3}\,f_2(x_{G^\prime},x_Q)\right)c_q^2\bigg)\,,
\end{align}
with $x_Q=m_Q^2/m_U^2$ and
\begin{align}
    f_1(x_V) = \frac{\log(x_V)}{x_V-1}\,,\qquad 
    f_2(x_V,x_Q) =\frac{1}{x_V-x_Q}\left(\frac{x_V\log(x_V)}{x_V-1}- \frac{x_Q\log(x_Q)}{x_Q-1} \right)\, .
\end{align}
\subsection{Expressions for $\delta R_{D^{(*)}}$ in terms of model parameters}

We report here the expressions for $\delta R_{D^{(*)}}$ in terms of the model parameters, in limit of up-alignment in the 12 quark sector. In this case, the Cabibbo angle comes from the down sector and,
as shown in Appendix~\ref{app:appendixA}, in the down-quark mass basis we have $\beta_{L}^{d\tau} \approx V_{cd} s_\chi s_q = V_{cd} \, \beta_{L}^{s\tau} / V_{cs} $. This implies 
\begin{equation}
A= \bigg(\frac{V_{cs}}{V_{cb}} \frac{\beta_{L}^{s\tau}}{\beta_{L}^{b\tau}}+\frac{V_{cd}}{V_{cb}} \frac{\beta_{L}^{d\tau}}{\beta_{L}^{b\tau}}\bigg) \approx \frac{V_{cs}}{V_{cb}} \frac{s_\chi}{c_\chi} \bigg(1+\frac{V_{cd}^2}{V_{cs}^2} \bigg) s_q  \approx \frac{s_q}{V_{cb}} \frac{s_\chi}{c_\chi} \big(1+\lambda^2\big) \,,
\end{equation}
where $\lambda=|V_{us}|=0.225$. Therefore, $A$ is a real parameter and we can write
\begin{align}
\delta R_{D} &\approx 2C_U \,\rho_{LL}^{\rm NLO} c_{\chi}^2  \bigg(1-1.50\,\eta_S\,\frac{\rho_{LR}^{\rm NLO}}{\rho_{LL}^{\rm NLO}}\,\frac{c_{R}}{c_\chi}\bigg) \,,\\
\delta R_{D^*} &\approx 2C_U\, \rho_{LL}^{\rm NLO} c_{\chi}^2   \bigg(1-0.12\,\eta_S\,\frac{\rho_{LR}^{\rm NLO}}{\rho_{LL}^{\rm NLO}}\,\frac{c_{R}}{c_\chi}\bigg) \,,
\end{align}
where we have defined $c_R = {\rm Re}(\beta_{R}^{b\tau})$. 
A constructive contribution to $\delta R_{D^{(*)}}$ requires $c_R < 0$. 
On the other hand, a too large (negative) $c_R$ would imply a large deviation between 
$\delta R_{D^*}$ and $\delta R_{D}$, which is not supported by present data. This motivates our choice of $c_R = -0.3$ for BP1.  Because $\delta R_{D^{(*)}}$ depends linearly on both $C_U$ and $s_q$, while $K-\bar K$ mixing scales as $s_q^4$, we work in the large $C_U$, small $s_q$ limit to maximize the contribution to $\delta R_{D^{(*)}}$ while passing the bounds on kaon mixing. Matching onto the simplified model of Ref.~\cite{Cornella:2021sby}, we identify
\begin{align}
C_U^{\rm simp} &= C_U  c_{\chi}^2\,,
\end{align}
which for $\chi \approx \pi /4$ implies $C_U \lsim 2(C_U^{\rm simp})_{\rm max} \approx 2\times 0.015 = 0.03$, where we have extracted $(C_U^{\rm simp})_{\rm max}$ from the left panel of Fig.~2.3 in Ref.~\cite{Cornella:2021sby}. This motivates the benchmarks $C_U = 0.015, \, 0.030$ considered in this work.

\section{Wilson coefficients}\label{app:appendixB}
In Table~\ref{tab:4QWCs} and Table~\ref{tab:SLWCs} we report the Wilson coefficients of operators contributing to $\Delta Q=2$ and $K\to \pi\nu\bar{\nu}$ transitions, as normalized in \eqref{eq:DeltaQ2_SMEFT} and \eqref{eq:sdnn_SMEFT}.

\begin{table}[t]
\begin{center}
\renewcommand{\arraystretch}{1.3}
\begin{tabular}{|c||c|c|c||c|c||c|c|c|}
\hline
 \multirow{3}{*} {}&\multicolumn{3}{c||}{Tree level }  	 &  \multicolumn{2}{c||}{Box diagrams } & 
 \multicolumn{3}{c|}{Flavor changing vertices }\\
 \multirow{2}{*} {}&\multicolumn{3}{c||}{$\cC_{qq}^{(a)ijkl}/\frac{1}{x_V}$}  	 &  \multicolumn{2}{c||}{$\cC_{qq}^{(a)ijkl}/ \frac{g_4^2}{32\pi^2}$} & 
 \multicolumn{3}{c|}{~~$\cC_{qq}^{(a)ijkl}/\left((\beta_L)_{bL}^{*} (\beta_L)_{sL} \frac{g_4^2 V_{V}}{32\pi^2 x_V}\right)$~~}\\
 \cline{2-9}

 & \multicolumn{2}{c|}{$\Gp$} &	$\Zp$ &  \multicolumn{2}{c||}{$[U_1~U_1]$ } &   \multicolumn{2}{c|}{$\Gp$} & $\Zp$ \\  
 \cline{1-9}
{~~$ijkl$~~} & $a=1$	& $a= 3 $	 &  $a=1$ & $a=1 $	 & $a=3$ & $~~~\,a=1\,~~~$ & $~~~\,a=3\,~~~$ & $a=1$\\
 \hline\hline
${3333}$  & $\frac{1}{12}$ &    $\frac{1}{4}$   &  $\frac{1}{24}$ & $\frac{1}{2}B_{qq}^{1111}$ &  $\frac{1}{2}B_{qq}^{1111}$ & $0$ & $0$  & $0$ \\[3pt]  \hline

${2333} $  & $0$ &    $0$   & $0$ & \phantom{aa}$- \frac{s_q}{2}\,B_{qq}^{1211}$\phantom{aa} &  $- \frac{s_q}{2} B_{qq}^{1211}$ & $ \frac{1}{12}$ & $ \frac{1}{4}$ & $ \frac{1}{24}$  \\[3pt]  \hline

${2233} $  & $-\frac{s_q^2-\epsilon_3}{6}$ &    $ 0 $   & $ \frac{s_q^2-\epsilon_1}{24 }$ & $\frac{s_q^2}{2}\,B_{qq}^{2211}$ &  $\frac{s_q^2}{2}\,B_{qq}^{2211}$ & $0$ & $0$  & $0$  \\[3pt]  \hline

${2332}$ & $\frac{s_q^2-\epsilon_3}{4}$  &   $\frac{s_q^2-\epsilon_3}{4}$  & $0$ & $\frac{s_q^2}{2}\,B_{qq}^{1212}$ & $\frac{s_q^2}{2}\,B_{qq}^{1212}$ & $0$ & $0$  & $0$ \\[3pt]  \hline

${2323}$ & $0$  &    $0$   & $0$ & $\frac{s_q^2}{2}\,B_{qq}^{1221}$ & $\frac{s_q^2}{2}\,B_{qq}^{1221}$ & $0$ & $0$  & $0$ \\[3pt]  \hline

${2223} $ & $0$  &    $0$   & $0$ & $-\frac{s_q^3}{2}\,B_{qq}^{2221}$ & $-\frac{s_q^3}{2}\,B_{qq}^{2221}$ & $\frac{s_q^2-\epsilon_3}{12} $ & $\frac{s_q^2-\epsilon_3}{4} $ & $\frac{s_q^2-\epsilon_1}{24} $  \\[3pt]  \hline

${2222}$ & $\frac{(s_q^2-\epsilon_3)^2}{12}$  &    $\frac{(s_q^2-\epsilon_3)^2}{4 }$  & $\frac{(s_q^2-\epsilon_1)^2}{24 }$ & $\frac{s_q^4}{2}\,B_{qq}^{2222}$ & $\frac{s_q^4}{2}\,B_{qq}^{2222}$  & $0$ & $0$ & $0$  \\[3pt]  \hline

${1133} $ & $\frac{\epsilon_3  }{6}$  &   $0$ & $-\frac{\epsilon_1}{24 }$ & $0$  & $0$ & $0$ & $0$ & $0$ \\[3pt]  \hline

${1331}$ & $-\frac{\epsilon_3  }{4}$  &   $-\frac{\epsilon_3  }{4}$ & $0$ & $0$  & $0$ & $0$ & $0$ & $0$ \\[3pt]  \hline

${1123}$ & $0$  &    $0$   & $0$ & $0$ & $0$ & $\frac{\epsilon_3}{6} $ & $0$ &  $-\frac{\epsilon_1}{24}  $  \\[3pt]  \hline

${1321} $ & $0$  &    $0$   & $0$ & $0$ & $0$ & $-\frac{\epsilon_3}{4} $ & $-\frac{\epsilon_3}{4} $ &  $0$  \\[3pt]  \hline

${1122} $ & $\frac{\epsilon_3(s_q^2-\epsilon_3)}{6}$  &    $0$   & $-\frac{\epsilon_1(s_q^2-\epsilon_1)}{24 }$ & $0$  & $0$  & $0$ & $0$ & $0$ \\[3pt]  \hline

${1221} $ & $-\frac{\epsilon_3(s_q^2-\epsilon_3)}{4}$  &    $-\frac{\epsilon_3(s_q^2-\epsilon_3)}{4}$   & $0$ & $0$  & $0$  & $0$ & $0$ & $0$ \\[3pt]  \hline

${1111} $  & $\frac{\epsilon_3^2}{12}$ &    $\frac{\epsilon_3^2}{4}$   & $\frac{\epsilon_1^2}{24 }$ & $0$ & $0$ & $0$ & $0$ & $0$ \\[3pt]  \hline

\end{tabular}
\caption{Coefficients $\cC_{qq}^{(a)ijkl}$ of the four-quark effective operators, normalized as in \eqref{eq:DeltaQ2_SMEFT}, at tree level and next-to-leading order.
The factors $x_V$ are $\xG$ and $\xZ$ for $G^{\prime}$ and $Z^{\prime}$ contributions.
The loop functions $B_{qq}^{ijkl}\equiv B_{qq}^{ijkl}(x_L)$, $V_{V}\equiv V_{\Gp}(x_L)$ and $V_{V}\equiv V_{\Zp_q}(x_L)$ for $\Gp$ and $\Zp$ contributions, can be found in \cite{Fuentes-Martin:2020hvc}. Note that $C_{qq}^{ijkl}=C_{qq}^{klij}$ and $C_{qq}^{jilk}=(C_{qq}^{ijkl})^*$.
\label{tab:4QWCs}} 
\end{center}
\end{table}

\begin{table}
\begin{center}
\renewcommand{\arraystretch}{1.3}
\begin{tabular}{|c||c||c||c|}
\hline
 \multirow{2}{*}{}& ~Tree level~   	 &  ~Box diagrams~  & Flavor changing vertices \\
 \multirow{2}{*} {}&$\cC_{\ell q}^{33ij}/\frac{1}{\xZ}$  	 &  $\cC_{\ell q}^{33ij}/ \frac{g_4^2}{16\pi^2}$ & 
 ~$\cC_{\ell q}^{33ij}/\left((\beta_L)_{bL}^{*} (\beta_L)_{sL} \frac{g_4^2 V_{\Zp_q}}{16\pi^2 x_V}\right)$~\\
\hline

 ~~$ij$~~ & $\Zp$ &  $[U_1~U_1]$  &   $\Zp$ \\  

 \hline\hline

$33 $  & $-\frac{1}{4}$ & $- B_{q\ell}^{1111}$ & $0$  \\[3pt]  \hline

$23 $  & $0$ & $ s_q B_{q\ell}^{1211}$ &  $-\frac{1}{8}$  \\[3pt]  \hline

$22 $  & $-\frac{s_q^2-\epsilon_1}{4}$ & $- s_q^2 B_{q\ell}^{2211}$ & $0$  \\[3pt]  \hline

$11$  & $\frac{\epsilon_1}{4}$ & $0$ &  $0$\\[3pt]  \hline
\end{tabular}
\caption{Coefficients $\cC_{\ell q}^{33ij}$ of the semileptonic effective operators relevant for $K\rightarrow \pi \nu\bar\nu$, normalized as in \eqref{eq:sdnn_SMEFT}, at tree level and next-to-leading order. The loop functions $B_{q \ell}^{ij\alpha\beta}\equiv B_{q \ell}^{ij\alpha\beta}(x_L,x_Q)$, and $V_{\Zp_q}\equiv V_{\Zp_q}(x_L)$ can be found in \cite{Fuentes-Martin:2020hvc}. Note that $C_{\ell q}^{33ji}=(C_{\ell q}^{33ij})^*$.
\label{tab:SLWCs}} 
\end{center}
\end{table}

\end{widetext}

\bibliographystyle{JHEP}
\bibliography{references}

\end{document}